\documentclass[fleqn,usenatbib]{mnras}

\usepackage{newtxtext,newtxmath}

\usepackage[T1]{fontenc}
\usepackage{ae,aecompl}
\usepackage[dvipsnames]{xcolor}
\usepackage[toc,page]{appendix}
\usepackage[dvipsnames]{xcolor}

\usepackage{graphicx}	
\usepackage{amsmath}	
\usepackage{booktabs}
\usepackage{longtable}
\usepackage{pdflscape}
\usepackage{soul}

\title[A New PULX Candidate in Cen A]{Significant or Not? The Impact of Randomisation During Data Reduction on Confirming a New Pulsating Ultraluminous X-ray Source Candidate in Centaurus A}

\author[A. H. Knight et al.]{
Amy H. Knight,$^{1}$\thanks{E-mail: amy.h.knight@durham.ac.uk}
Timothy P. Roberts,$^{1}$
Callum Potter,$^{1}$ 
Alistair T. Pagan,$^{1}$
and Dominic J. Walton$^{2}$
\\
$^{1}$ Centre for Extragalactic Astronomy, Department of Physics, Durham University, South Road, Durham DH1 3LE, UK \\
$^{2}$ Centre for Astrophysics Research, University of Hertfordshire, College Lane, Hatfield, AL10 9AB, UK\\
}

\date{Accepted XXX. Received YYY; in original form ZZZ}

\pubyear{2025}

\begin{document}
\label{firstpage}
\pagerange{\pageref{firstpage}--\pageref{lastpage}}
\maketitle

\begin{abstract}
We report the discovery of a new candidate pulsating ultraluminous X-ray source (PULX) in NGC 5128 (Centaurus A). The candidate, 4XMM J132542.2--425943, is a transient source, identifiable as a clear X-ray point source for $\sim 8$ months in 2014, during its only major recorded outburst. The source flux exceeded $10^{-12}$ erg cm$^{-2}$ s$^{-1}$ at the peak of the outburst. The long-term light curve of 4XMM J132542.2--425943 shows two further, less luminous detections in 2017 and 2024, but was otherwise in quiescence. This behaviour is similar to the class of pulsating transients with outbursts that reach the ultraluminous regime, which includes the well-studied Galactic PULX, \textit{Swift} J0243.6+6124. However, 4XMM J132542.2--425943 displays a soft X-ray spectrum, making this source distinct from the existing population of PULXs, which typically show hard spectra below $10$ keV. We searched the 2014 \textit{XMM-Newton} observations for X-ray pulsations, revealing coherent, sinusoidal X-ray pulsations at a frequency of $1.27$ Hz in one \textit{XMM-Newton} observation (ObsID 0724060801), at a marginal significance. For this signal we measure a pulsed fraction, PF$~\approx~15 - 17~\%$ and $\dot{f}~\sim~4~\times~10^{-9}$~Hz~s$^{-1}$. However, we find that the intrinsic randomisation employed by \textit{XMM-Newton's} Science Analysis Software, \textsc{XMM-SAS}, during the data reduction procedure introduces considerable uncertainty in the strength of our marginal pulsations, which varies significantly between consecutive data reduction iterations. We explore the impact of this randomisation and demonstrate that it can generate widespread false positives and false negatives, which, in the context of PULX searches, may cause viable candidates to be unnecessarily discarded or vice versa.
\end{abstract}



\section{Introduction}
Ultraluminous X-ray sources (ULXs) are accreting, usually extragalactic, X-ray sources located away from the nuclear regions of their host galaxies, with luminosities that match or exceed the nominal Eddington limit for stellar-mass objects, $\sim 10^{39}$ erg s$^{-1}$ (see e.g. \citealt{Kaaret2017, King2023} for reviews). Initial explanations for the high X-ray luminosities of ULXs focused either on sub-Eddington accretion onto intermediate-mass black holes (IMBHs, $10^{3} M_{\odot} \leq M_{\rm{BH}} \leq 10^{5} M_{\odot}$, e.g. \citealt{Colbert1999, Strohmayer2009, Sutton2012}) or near- and/or super-Eddington accretion onto stellar mass black holes ($\leq 100 ~M_\odot$, e.g. \citealt{Stobbart2006, Zampieri2009}). However, the discoveries of coherent X-ray pulsations from several known ULXs (see \citealt{Bachetti2014, Furst2016, Israel2017, Carpano2018, Sathyaprakash2019, RodriguezCastillio2020, Ducci2025} and potentially \citealt{Pintore2025} for examples) have provided unwavering evidence that some ULXs host neutron stars (NSs) and proof that accretion rates in ULXs can far exceed the Eddington limit, in some cases by apparent factors of hundreds \citep{Israel2017}. 

Pulsating ULXs (hereafter PULXs) are challenging to reconcile with sub-Eddington accretion models, such as those used to describe accretion in Galactic NS X-ray binaries. Reconciliation may be possible under certain circumstances if the NS has a strong magnetic field, for example ($> 10^{13}$ G, see e.g., \citealt{DallOsso2015}). Indeed, the presence of a strongly magnetised NS likely creates an optically-thick accretion curtain within the magnetosphere of the ULX, the precession of which with the NS spin creates the sinusoidal pulsations inherent to PULXs \citep{Mushtukov2017, Conforti2025}. However, a possible electron cyclotron resonance scattering feature (CRSF) in a ULX in M51 implies a lower magnetic field strength ($\sim 10^{12}$ G; \citealt{Middleton2019}. See also \citealt{Brightman2018} for the original detection of this feature). NSs with lesser magnetic field strengths may accrete supercritically in a similar fashion to that proposed for black holes (\citealt{Kluzniak2015}; see \citealt{Shakura1973, Poutanen2007} for a description of supercritical accretion), such that their extreme luminosities may largely be the result of strong geometric beaming \citep{King2009}. This picture would be consistent with the large body of observational results for ULXs that support supercritical accretion, including the X-ray spectra \citep{Gladstone2009}, correlated spectral and temporal variability \citep{Sutton2013} and the detection of the signatures of massive outflowing winds \citep{Pinto2016}. However, the detection of an ultrafast outflow from a PULX \citep{Kosec2018} has led to the development of hybrid models that incorporate both magnetic NSs and features of a supercritical flow \citep{Mushtukov2019}, but the balance between the degree of beaming and the magnetic field strength remains an open question \citep{Kovlakas2025}. Indeed, some intriguing recent results indicate that the anisotropy predicted by supercritical models may not be the key driver of the observed properties of ULXs \citep{Gurpide2024}. Notably, the apparent change in the orbital period of M82 X-2 \citep{Bachetti2022}, which provides a constraint on the mass transfer rate within the binary, does not imply strong beaming despite an apparent luminosity $\sim 100 L_{\rm{edd}}$.

Clearly, a better understanding of the physics of accretion in PULXs in particular, and ULXs hosting NSs in general, is required. One way to do this is to widen the sample of PULXs we have available to study. There are suggestions that a significant fraction of ULXs may host accreting NSs rather than black holes, \citep{Pintore2017, Koliopanos2017, Walton2018}. However, confidently identifying a NS accretor is challenging. Unambiguous confirmation of a NS via the detection of pulsations usually requires relatively rich datasets (many thousands of X-ray counts) to overcome the low pulse fractions, which are only available for a few tens of ULXs at most (cf. ULX catalogues such as \citealt{Walton2022} and \citealt{Tranin2024}). The difficulty in identifying pulsations is heightened by the characteristics of the pulsations themselves, as they are often transient (c.f. \citealt{Fuerst2023}). There is also uncertainty as to whether most NS ULXs will even show pulsations \citep{Middleton2017}. Given these difficulties, it is natural to focus searches for new PULXs on the most promising candidates. There are various suggestions for the best candidates, including those ULXs with spectral and temporal properties most closely matching the known PULXs \citep{Gurpide2021}; ULXs showing the highest amplitudes of variability either as a result of the propeller effect \citep{Earnshaw2018, Song2020} or as a result of the precession of the narrow funnel structures that geometrically beam the emission from NS ULXs \citep{Khan2022}; or simply the more luminous examples of the population \citep{Roberts2023}.

Here we report a result obtained through a different approach: a thorough reanalysis of available archival data, utilising one of the most up-to-date source catalogues \citep{Walton2022} as the basis for the source selection. Using this approach, we reveal a new candidate PULX by finding a plausible pulsation signal associated with 4XMM J132542.2--425943 (hereafter J1325), a hitherto unstudied transient ULX located in NGC 5128 (Centaurus A; see catalogue entries from \citealt{Walton2022, Tranin2024}). We structure this paper as follows. In Section \ref{sx:Data}, we describe our data selection and reduction procedures, and then explore the spectral and temporal properties of J1325 in Section \ref{sx:properties}. We detail our novel two-step approach to pulsation searches in Section \ref{sx:Pulse}, which includes a thorough exploration of signal strength variations arising from the data reduction procedure. We discuss our results in Section \ref{sx:Discuss} and conclude in Section \ref{sx:conc}.

\vspace*{-0.3cm}
\section{Source Selection and Data Reduction}
\label{sx:Data}

\begin{table}
    \centering
    \begin{tabular}{cccc}
    \multicolumn{4}{| c |}{\textit{XMM-Newton}}\\
    \hline
    ObsID & Start Date \& Time & Duration & Source Detection \\
    & [UTC] & [s] & Significance \\
    \hline
    \vspace*{0.1cm}
    0724060701 & 2014-01-06 19:00:27 & 26500 & $97.7\sigma$ \\
    \vspace*{0.3cm}
    0724060801 & 2014-02-09 15:53:25 & 23400 & $121.4\sigma$ \\ 

    \multicolumn{4}{| c |}{\textit{Chandra}}\\
    \hline
    \vspace*{0.1cm}
    16276 & 2014-04-24 15:42:57 & 5090 & $100.2\sigma$ \\
    \vspace*{0.1cm}
    16277 & 2014-09-08 04:47:53 & 5080 & $46.6\sigma$ \\ 
    \vspace*{0.3cm}
    19747 & 2017-05-15 21:39:06	 & 5070 & $4.82\sigma$ \\ 

    \multicolumn{4}{| c |}{\textit{Swift-XRT}}\\
    \hline
    \vspace*{0.1cm}
    00049671009 & 2024-02-02 15:57:57 & 2192 & $3.25\sigma$ \\
    
    \hline
    \end{tabular}
    \vspace*{-0.2cm}
    \caption{A summary of the \textit{XMM-Newton}, \textit{Chandra} and \textit{Swift-XRT} observations in which 4XMM J132542.2--425943, is identifiable as an X-ray point source. The significances reported for \textit{XMM-Newton} are calculated from EPIC-PN.}
    \label{tb:obs}
\end{table}

We utilised the \citet{Walton2022} multi-mission catalogue of ULX candidates (hereafter W22) to identify suitable candidates for pulsation searches. W22 consists of 1843 ULX candidates across 951 host galaxies, compiled from data releases of \textit{XMM-Newton} (4XMM-DR10), \textit{Swift} (2SXPS) and \textit{Chandra} (CSC2.0), after the removal of known contaminants. We first filtered the W22 catalogue for sources with count-rich datasets, high observed fluxes and observations by \textit{XMM-Newton's} EPIC-PN detector to ensure the data had sufficient time resolution for pulsation searches. We then discarded any observations of known PULXs, leaving 44 observations of 19 different ULXs. 

For each of these 44 observations, we performed a single data reduction of the \textit{XMM-Newton} Observation Data Files (ODFs), following the standard reduction procedure (keeping all arguments to their default values) and conducted an initial analysis. The initial analysis consisted of a single $0.01 - 5.0$~Hz Fourier domain acceleration search in the $1 - 10$ keV energy band, using the \texttt{accelsearch} procedure within \textsc{hendrics} \citep{Bachetti2018}, with all other parameters kept to their default values. If this acceleration search returned a pulsation candidate with a power greater than $35$, we subsequently performed a Z$^{2}_{3}$ search in a $0.1$ Hz band centred on that candidate, again, keeping all other parameters at their default values. If the results were consistent, we then repeated the two searches in the $0.2 - 10$ keV band to see if the pulsation signal appeared stronger in the harder band, as is typical for PULXs. This initial analysis served as our source selection procedure. By looking for sources that returned consistent pulsation candidates across the searches, we could select the most promising PULX candidates to study further. 

From this initial analysis, J1325, whose host galaxy is NGC 5128 (Centaurus A), stood out because the searches all returned the same candidate frequency of $1.27$ Hz. This preliminary candidate appeared in a single \textit{XMM-Newton} ObsID (0724060801) and appeared stronger when running the initial analysis in the $1 - 10$ keV band than the $0.2 - 10$ keV band. Furthermore, there is a lack of a previously published analysis on J1325. Therefore, we chose to focus on this source and used all publicly available archival observations of NGC 5128 from \textit{XMM-Newton} and \textit{Chandra} to characterise the source. We describe their respective data reduction procedures in the remainder of this Section. We further checked the public data archives to see if another instrument with suitable time resolution for pulsations searches (e.g. \textit{NICER}, \textit{NuStar}) had captured J1325 during an outburst, but do not find any other suitable observations. Therefore, we focus on \textit{XMM-Newton} to conduct our thorough pulsation analysis and note that a future observation of J1325 during an outburst, by an instrument with sufficient temporal resolution will be necessary to confirm our findings. However, we did identify archival exposures of NGC 5128 taken by the Neil Gehrels Swift Observatory's (hereafter \textit{Swift}) X-ray Telescope (XRT) and incorporate these where appropriate.

\subsection{\em Chandra}
\label{sx:chandra}

\begin{figure}
\includegraphics[width=\columnwidth]{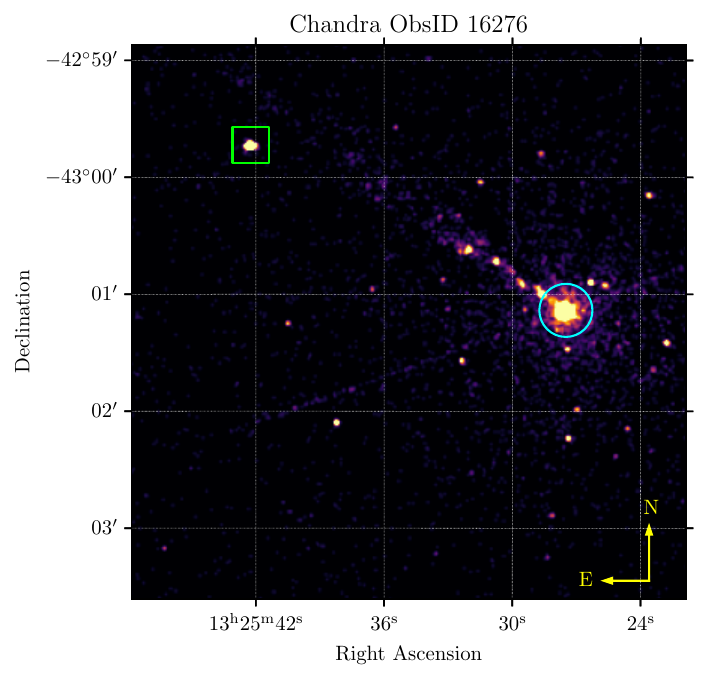} 
\vspace*{-0.65cm}
\caption{The field of view of \textit{Chandra} ObsID 16276 in which our PULX candidate, 4XMM J132542.2--425943, is identifiable as an X-ray point source. The PULX candidate lies close to the axis of the jet emitted from the active nucleus of Centaurus A (South-West of 4XMM J132542.2--425943 in this image, circle).}
\label{fig:ChandraObs}
\end{figure}

We reduced all $71$ archival \textit{Chandra} observations of NGC 5128 with the Chandra Interactive Analysis of Observations software, \textsc{ciao}, version 4.17 \citep{Fruscione2006} in conjunction with \textsc{caldb} version 4.11.6. We reprocessed the data using \texttt{chandra$\_$repro}, keeping all arguments to their default values. For each observation, we viewed the resulting level=2 event file with \textsc{ds9}, determining that the position of J1325 falls within the field of view of the detector in $52$ of the $71$ exposures. Among these, we detect J1325 in 3 observations: ObsIDs 16276, 16277 and 19747. Details of these three detections are provided in Table \ref{tb:obs}. We further show the position of J1325 in ObsID 16276 (square; RA: 13h:25m:42.2s, DEC: -42d:49m:43.0s) with respect to the active nucleus, Cen A (circle) in Figure \ref{fig:ChandraObs}. Here we see J1325 is positioned close to the axis of the jet emitted from Cen A.

Of the three observations in which J1325 is identifiable as an X-ray point source, ObsID 16276, is the only exposure with sufficient counts to perform spectral analysis. In this case, we extracted a $0.5 - 7$ keV spectrum by executing \texttt{specextract}. We applied a $5$ arcsecond radius circular source region centred on the source coordinates and one of twice this radius at a position close to J1325 that avoided jet emission to model the background. We set \texttt{weight=no} to create an Auxiliary Response File (ARF) for a point-like source and applied energy-dependent point-source aperture corrections using \texttt{arfcorr} with \texttt{correctpsf=yes}. Finally, we grouped the resulting background-subtracted source spectrum to have at least 25 counts per bin to enable the use of chi-squared statistics. This spectrum is analysed in Section \ref{sx:Chan_Spec}

\vspace*{-0.25cm}
\subsection{\em XMM-Newton}
All seven publicly available archival observations of NGC 5128 were obtained from the \textit{XMM-Newton} Science Archive and reduced using the \textit{XMM-Newton} Science Analysis Software, \textsc{XMM-SAS} version 21.0.0 and the latest calibration files (\citealt{SAS}; see also \citealt{SAS2025} for an up-to-date description of \textsc{XMM-SAS}). We began by reprocessing the ODFs to obtain calibrated and concatenated EPIC-PN and EPIC-MOS event lists for each observation using the \textsc{XMM-SAS} tasks \texttt{epproc} and \texttt{emproc}, respectively, with default values. To identify and exclude periods of enhanced background, we extracted high-energy light curves using \texttt{evselect}, specifying single patterns and energies between 10 and 12 keV for EPIC-PN and energies greater than 10 keV for EPIC-MOS. Subsequently, we used \texttt{tabgtigen} to determine the good-time intervals (GTIs) for each observation using a rate expression set by identifying a count rate threshold just above the background level from a stable 5~ks portion of the high-energy light curve \footnote{\url{https://www.cosmos.esa.int/web/xmm-newton/sas-thread-epic-filterbackground-in-python}}. These thresholds were determined by adding $0.05$~ct/s to the average count rate in the selected 5~ks segment of the high-energy light curve. For these data, the count rate threshold is typically $0.3 - 0.4$ cts/s for EPIC-PN and $0.2 - 0.3$ cts/s for EPIC-MOS. Filtered event lists were then created by applying the GTI filters with \texttt{evselect} and were corrected to Barycentric Dynamical Time (DE200 ephemeris) using \texttt{barycen}. We subsequently generated an image of each observation from the clean, barycentric EPIC-PN event lists. From these, we identified a bright X-ray point source at the coordinates of J1325 (RA: 13h:25m:42.2s, DEC: -42d:49m:43.0s) in two of the seven observations, ObsIDs 0724060701 and 0724060801 (both in small window mode; see also Table \ref{tb:obs}). We applied 20 arcsecond radius circular source regions centred on the source coordinates to extract source event lists in the $0.2-10.0$ keV and $1.0-10.0$ keV bands for all three EPIC detectors. The pulsation searches described in Section \ref{sx:Pulse} are conducted using these EPIC-PN source event lists.

We followed the standard spectrum extraction procedure using \textsc{XMM-SAS}\footnote{\url{https://www.cosmos.esa.int/web/xmm-newton/sas-thread-mos-spectrum}} to extract spectra from ObsIDs 0724060701 and 0724060801. We first extracted a source plus background spectra from each of the EPIC detectors in the $0.2-10.0$ keV band using \texttt{evselect}, specifying all channels (0-11999 for both MOS detectors and 0-20479 for PN) and a spectral bin size of 5. We used singles, doubles, triples and quadruples when extracting MOS1 and MOS2 spectra and singles and doubles when extracting the PN spectra. We repeated this step to extract background spectra for subtraction, utilising 20 arcsecond source-free regions close to the position of J1325 for the background subtraction, while avoiding emission from the X-ray bright jets of NGC 5128's AGN that the ULX position is projected close to (see Section \ref{sx:chandra} and Figure \ref{fig:ChandraObs} for details). While it is typical to use a background region of at least twice the radius of the source region, we note that this was not possible, as we required the background region to avoid the jet emission and be positioned on the same part of the detector as the source. We also note that ObsIDs 0724060701 and 0724060801 were taken in small-window mode, which limits the available detector area from which to obtain a background. We applied \texttt{backscale} to determine the areas of the source and background regions used to extract the spectral files, accounting for any bad pixel locations, and then generated redistribution matrices and ancillary response files using \texttt{rmfgen} and \texttt{arfgen}, respectively. Finally, we grouped the spectra to have at least 25 counts for each spectral channel and did not oversample the intrinsic energy resolution by a factor higher than 3. These spectra are studied in Section \ref{sx:xmmspec}. 

\vspace*{-0.6cm}
\section{Source Properties}
\label{sx:properties}
\begin{figure*}
    \centering
    \includegraphics[width=\textwidth]{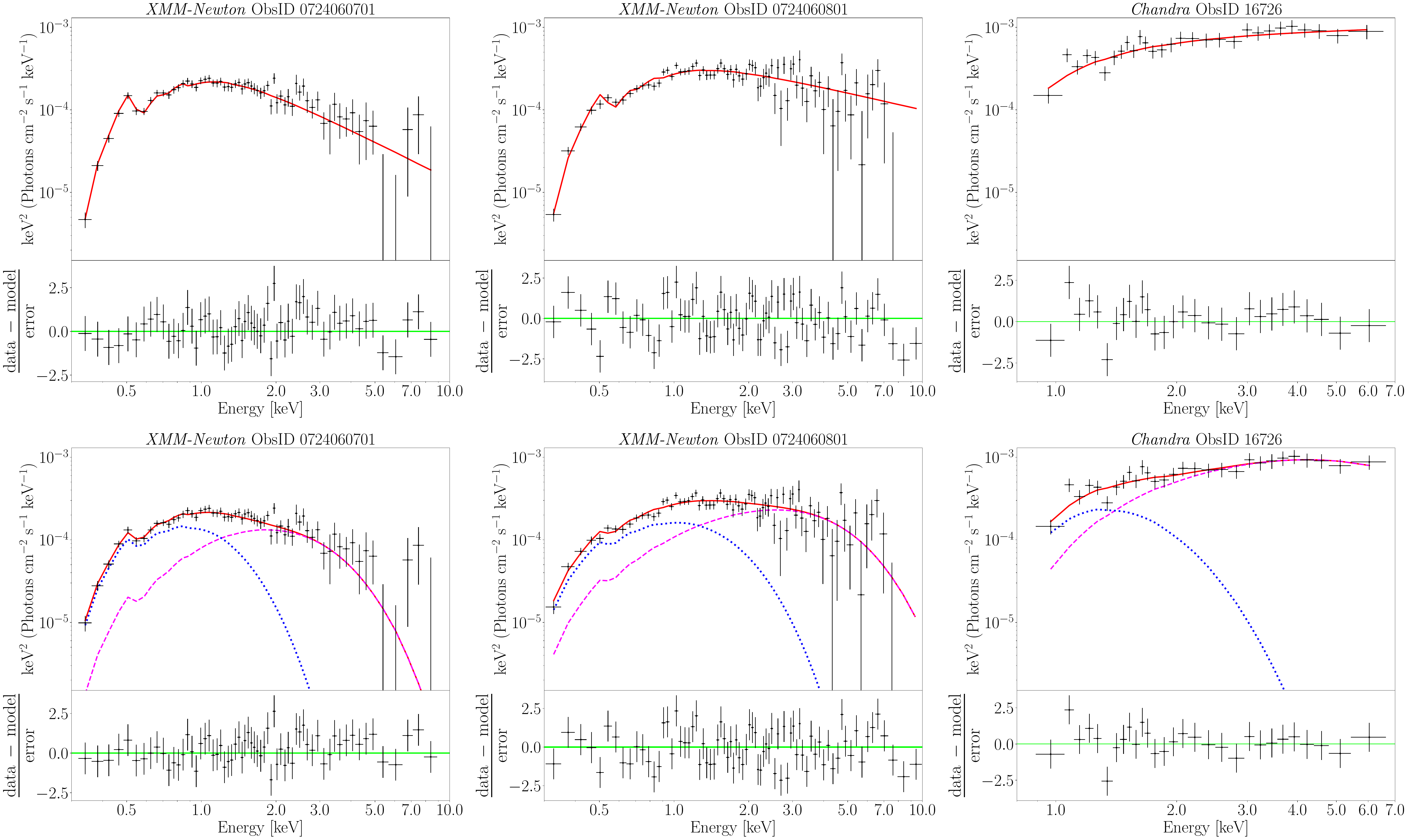}
    \vspace*{-0.5cm}
    \caption{Unfolded \textit{XMM-Newton} ObsIDs 0724060701 (left) and 0724060801 (middle) and \textit{Chandra} ObsID 16276 (right) spectra fit with the single-component model \texttt{tbabs(tbabs(powerlaw))} (top row) and the two-component model \texttt{tbabs(tbabs(diskbb+diskbb))} (bottom row). In each panel, the best-fitting total model is shown by the solid line and the residuals are shown in the lower segment. In the bottom row, the magenta dashed and blue dotted lines represent two \texttt{diskbb} model components. Note the \textit{XMM-Newton} panels show only the data and residuals from the EPIC-PN detector, but the plotted models were obtained through simultaneous spectral fitting to the data from all three EPIC detectors.}
    \label{fig:Spectra}
\end{figure*}

\subsection{{\em{XMM-Newton}} Spectral Fitting and Flux Measurements}
\label{sx:xmmspec}

For \textit{XMM-Newton} ObsIDs 0724060701 and 0724060801, we fit the extracted MOS1, MOS2 and PN spectra simultaneously using \textsc{xspec} version 12.13.1, and try four different models. In all four models, the first \texttt{tbabs} component is fixed to a value of $8.41 \times 10^{20}$ atom cm$^{-2}$, as determined by the Colden Galactic Neutral Hydrogen Density Calculator~\footnote{\url{https://cxc.harvard.edu/toolkit/colden.jsp}} using the NRAO dataset \citep{Dickey1990}. We also use a constant factor (fixed to a value of 1.0 for the PN spectrum and allowed to vary for the MOS1 and MOS2 spectra) to model the differences in detector calibration, finding that they are consistent to within a few per cent. 

We begin with an absorbed power law model, \texttt{tbabs(tbabs(powerlaw))}, and show the resulting fits in the top row of Figure \ref{fig:Spectra} (left and middle panels, respectively, for ObsIDs 0724060701 and 0724060801). The corresponding best-fitting parameters are presented in Table \ref{tb:single}, where we see that the respective best-fitting power law indices are $\sim 3.5$ and $\sim 2.7$, indicating that J1325 was in a soft spectral state, which is atypical for PULXs. For simplicity, we utilise the absorbed power law model to determine the observed $0.5-10$~keV source flux in ObsIDs 0724060701 and 0724060801, and convert these into luminosities in Table \ref{tb:single}, assuming a distance of $3.8$ Mpc \citep{Harris2010}. For ObsIDs 0724060701 and 0724060801 we obtain $(9.24 \pm 0.01) \times 10^{38}$ erg s$^{-1}$ and $(1.70 \pm 0.02) \times 10^{39}$ erg s$^{-1}$, consistent with the ULX luminosity regime. These measurements appear as blue squares in Figure \ref{fig:Long_Term_Lightcurve}. We then employ an absorbed disc blackbody model, \texttt{tbabs(tbabs(diskbb))}. Here, the best-fitting disc temperatures are $\sim 0.4$ and $\sim 0.6$ keV, respectively, for ObsIDs 0724060701 and 0724060801. For the most luminous ULXs, adequate spectral modelling below 10~keV typically requires two thermal components (see below). In these cases, the softer thermal components typically have temperatures between 0.2 and 0.5 keV, which roughly align with our single-component measurements. However, when comparing these two single-component models in Table \ref{tb:single}, we statistically favour the absorbed power law model, which achieves $\chi^{2} = 159.77$ and a null-hypothesis probability of $0.240$ for $\nu = 148$ and $\chi^{2} = 170.71$ and a null-hypothesis probability of $0.449$ for $\nu = 169$, respectively for ObsIDs 0724060701 and 0724060801. In contrast, we statistically reject the absorbed disc blackbody model which achieves $\chi^{2} = 219.15$ and a null-hypothesis probability of $1.31 \times 10^{-4}$ for $\nu = 148$ and $\chi^{2} = 273.06$ and a null-hypothesis probability of $ 6.90 \times 10^{-7}$ for $\nu = 169$ for ObsIDs 0724060701 and 0724060801, respectively.

As noted above, many good quality ULX spectra taken with \textit{XMM-Newton} or \textit{Chandra} are better described by a two-component model, consisting of either an accretion disc blackbody and a power law, as is commonly used to describe Galactic X-ray binary spectra, or two blackbody components. Therefore, we try to improve upon the absorbed disc blackbody fits by adding a second component. We begin with the model \texttt{tbabs(tbabs(diskbb+powerlaw))}. When fitting ULX spectra below $10$ keV with a model of this form, the blackbody temperature is typically cool, $\sim 0.1-0.3$ keV. This cool blackbody represents a soft excess to a dominant power law component, consistent with the soft and hard ultraluminous regimes detailed in \citep{Sutton2013}. Here, a power law with $\Gamma > 2$ distinguishes a soft from a hard ultraluminous spectrum. ObsIDs 0724060701 is fitted best with a cool blackbody temperature of $\sim 0.150$ keV and a very soft photon power law index of $\Gamma \sim 3.41$, consistent with a soft ultraluminous spectrum. However, ObsID 0724060801 is different, with a harder best-fitting photon power law index of $\Gamma \sim 2.80$ and a hotter blackbody with a temperature of $\sim 0.862$ keV. Despite the difference to ObsIDs 0724060701, these spectral parameters still allow it to be classed as a soft ultraluminous spectrum, especially given the dominant flux of the power law component (see figure 2 of \citealt{Sutton2013}). As in Table \ref{tb:double}, both fits are statistically valid, with null-hypothesis probabilities of $0.239$ and $0.524$ for ObsIDs 0724060701 and 0724060801, respectively.

The use of two blackbody components when modelling the spectra of ULXs arose after the discovery that the hard emission does not always present as a power law. It instead shows curvature in the $2-10$ keV band (e.g. \citealt{Roberts2005, Feng2005}) and appears statistically significant in the majority of high signal-to-noise cases (e.g. \citealt{Stobbart2006, Gladstone2009}). This curvature has since been unambiguously confirmed by NuSTAR (e.g. \citealt{Bachetti2013, Walton2014, Rana2015}). As such, we also implement the model \texttt{tbabs(tbabs(diskbb+diskbb))} and show the resulting fits in Figure \ref{fig:Spectra} (bottom left and middle, respectively for ObsIDs 0724060701 and 0724060801). As shown in Table \ref{tb:double}, our fits comprise a hotter thermal component, $\gtrsim 0.69$ keV, which describes the harder part of the spectra, and a cooler thermal component, $0.2 - 0.3$ keV, that represents the softer part. These parameters are generally consistent with the population of ULXs fit with similar two-thermal component models, albeit at the cooler end of the distribution for the hard component (cf. \citealt{Gurpide2021} Table 2). Overall, using a model with two thermal components rather than one improves the fits (see Tables \ref{tb:single} and \ref{tb:double}). The corresponding fit statistics with two thermal components are $\chi^{2} = 147.11$ and a null-hypothesis probability of $0.459$ for $\nu = 146$ and $\chi^{2} = 152.63$ and a null-hypothesis probability of $0.785$ for $\nu = 167$, respectively, for ObsIDs 0724060701 and 0724060801.

Finally, we consider the remaining five \textit{XMM-Newton} observations of NGC 5128, reduced in Section \ref{sx:Data}. Of these, we discard ObsIDs 0724060501 and 0724060601, as the source coordinates fall outside the detector's field-of-view. For the other three observations, ObsIDs 0093650201, 0093650301 and 0863890201, we utilised 20 arcsec circular source regions, centred on the source coordinates, to place $3~\sigma$ upper limits on the source flux. In each case, the upper limit on the source count rates were converted to a $0.5 - 10$ keV flux, assuming a photon power law index of $\Gamma = 3.5$, a galactic absorption of $N_{\rm{H}}= 8.41 \times 10^{20}$ atom cm$^{-2}$ and a local absorption of N$_{\rm{H}} = 2.19 \times 10^{21}$ atom cm$^{-2}$ (consistent with the best-fitting spectral model of the fainter \textit{XMM-Newton} detection ObsIDs 0724060701; see Table \ref{tb:single}). These limits appear as blue arrows in Figure \ref{fig:Long_Term_Lightcurve}.

\subsection{{\em{Chandra}} Spectral Fitting and Flux Measurements}
\label{sx:Chan_Spec}

For all \textit{Chandra} observations containing the source coordinates, we determine the flux or an upper limit on the flux in the $0.5 - 7.0$ keV band for ACIS and $0.1 - 10.0$ keV for HRC using the \textsc{ciao} routine, \texttt{srcflux}. For each observation, we specified the source RA and DEC as 13h:25m:42.2s, -42d:49m:43.0s, and allowed \texttt{srcflux} to determine the source and background regions, which by default are circular source regions centred on the given coordinates and the radius is set as the size of a circle needed to enclose 90~$\%$ of the PSF at 1.0~keV. We also supplied \texttt{srcflux} with a spectral model with a photon power law index of $\Gamma = 3.5$, a Galactic absorption of $N_{\rm{H}}= 8.41 \times 10^{20}$ atom cm$^{-2}$ and an additional absorption of $N_{\rm{H}} = 2.19 \times 10^{21}$ atom cm$^{-2}$, consistent with the best-fitting spectral model of the fainter \textit{XMM-Newton} detection ObsIDs 0724060701 (see Section \ref{sx:xmmspec} and Table \ref{tb:single}). We removed ObsIDs 26039 and 7797 from this analysis because the position of J1325 is coincident with the out-of-time events, resulting in a false positive source detection in both cases. For all other \textit{Chandra} exposures containing the source coordinates, the measured fluxes or upper limits are corrected to the $0.5 - 10$~keV \textit{XMM-Newton} bandpass and shown in Figure \ref{fig:Long_Term_Lightcurve} with red circles and arrows, respectively.

The exception to this is ObsID 16276, which is the only \textit{Chandra} observation with sufficient counts to enable spectral analysis of J1325. In this case, we fit the extracted $0.5 - 7.0$ keV spectrum with \textsc{xspec} using the same four models applied to the \textit{XMM-Newton} spectra in Section \ref{sx:xmmspec}. The results are presented alongside the \textit{XMM-Newton} best fits in Tables \ref{tb:single} and \ref{tb:double}, and Figure \ref{fig:Spectra}. Despite the spectrum having fewer counts than those extracted from the two \textit{XMM-Newton} observations, we chose to include ObsID 16276 because it provides the best representation of the spectral state of J1325 near the peak of the 2014 outburst. This is evident from the observed, bandpass- corrected (to the $0.5 - 10$ keV band) flux measurement in Table \ref{tb:single} and Figure \ref{fig:Long_Term_Lightcurve}, determined from the best-fitting absorbed power-law model. The observed flux is subsequently converted to a luminosity of $4.83 \times 10^{39}$ erg s$^{-1}$, assuming a distance of 3.8 Mpc \citep{Harris2010}. However, as is apparent from Tables \ref{tb:single} and \ref{tb:double}, the lower quality spectrum prevents us from statistically separating the four fits as the null-hypothesis probabilities for models 1-4 are $0.708$, $0.593$, $0.589$ and $0.572$, respectively. Nonetheless, when comparing the spectral fits across the three observations (January, February and April 2014), we find evidence that the source hardens as it approaches the peak of the outburst. This trend is most apparent in Table \ref{tb:single}, which shows the spectral fitting results using an absorbed power law model. Here, we see the best-fitting power law indices decrease as the outburst evolves. Similarly, when using a single-component black body model, the black body temperature increases as the outburst evolves (see Table \ref{tb:single}), likely due to the source increasing in luminosity as it approaches the peak of the outburst.

Overall, our spectral analysis shows a key result -- our PULX candidate was in a soft spectral state during the observation in which we identified coherent X-ray pulsations (\textit{XMM-Newton} ObsID 0724060801; see Section \ref{sx:Data} for the preliminary detection and Section \ref{sx:Pulse} for detailed analysis of the pulsations). The apparent spectral softness of J1325 makes it an outlier from the existing population of PULXs, which are known to exhibit harder-than-average X-ray spectra below $10$ keV (cf. \citealt{Gurpide2021}). This finding admits two interpretations: (i) J1325 may be the first reported spectrally soft PULX, implying a gap in our current understanding of the pulsation mechanism in ULXs; or (ii) it is not a PULX, and the X-ray pulsations we marginally detect do not originate from J1325. We discuss these interpretations further in Section \ref{sx:Discuss}.

\subsection{Long Term Light Curve}

\begin{figure*}
    \centering
    \includegraphics[width=\textwidth]{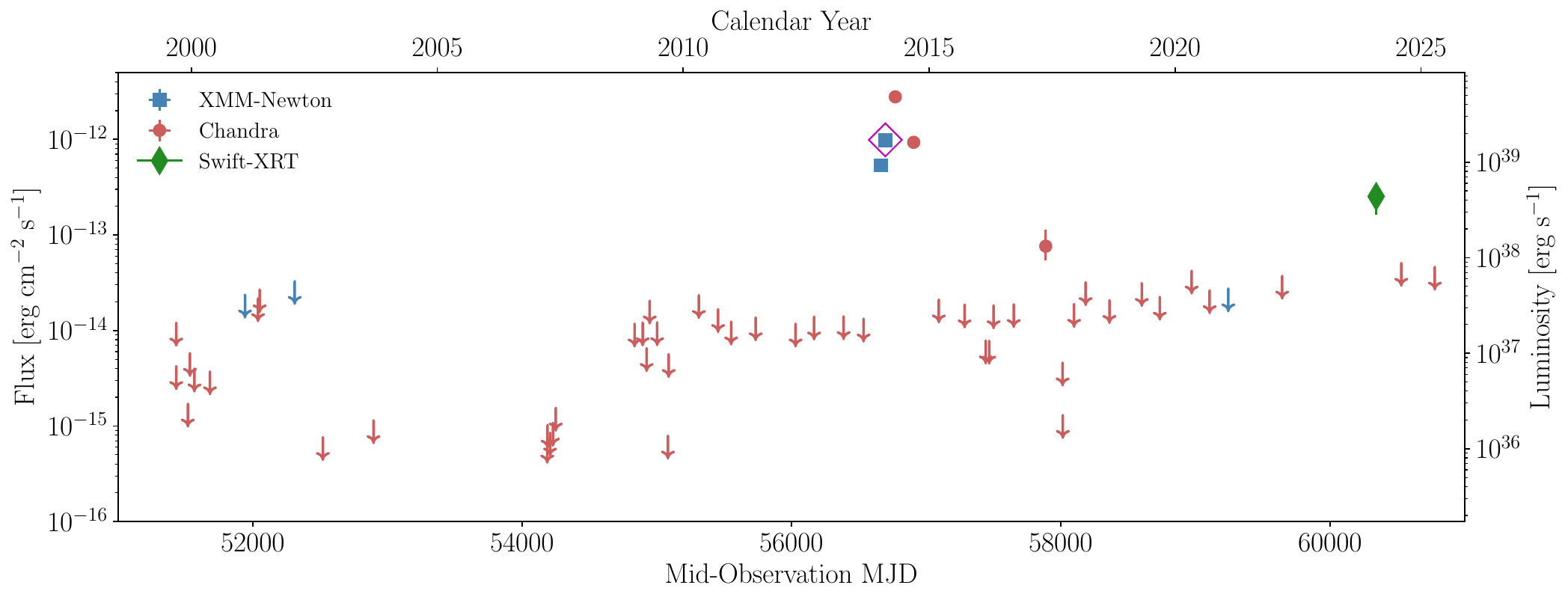}
    \vspace{-0.65cm}
    \caption{Long term light curve of the candidate pulsating ultraluminous X-ray source, 4XMM J132542.2--425943. The source has undergone one bright outburst between January - September 2014, detected by both \textit{XMM-Newton} (squares) and \textit{Chandra} (circles). 4XMM J132542.2--425943 briefly brightened again in May 2017 and February 2024, the latter of which was captured by \textit{Swift-XRT} (diamond). The arrows represent $3~\sigma$ upper limits on the flux of 4XMM J132542.2--425943, converted using a photon power law index of $\Gamma = 3.5$, a Galactic absorption of $N_{\rm{H}}=$ 8.41 $\times 10^{20}$ atom cm$^{-2}$ and an additional absorption of $N_{\rm{H}} = 0.219 \times 10^{22}$ atom cm$^{-2}$ (consistent with the best-fitting spectral model of the fainter \textit{XMM-Newton} detection; see Section \ref{sx:properties}). Corresponding luminosities are determined assuming a distance of $3.8$ Mpc \citep{Harris2010}. The magenta diamond highlights the observation in which we detect marginal X-ray pulsations at $1.27$~Hz.}
    \label{fig:Long_Term_Lightcurve}
\end{figure*}

In Figure \ref{fig:Long_Term_Lightcurve}, we compile the above flux measurements and limits into a long-term light curve of J1325. Here, we see that the two 2014 \textit{Chandra} observations, ObsIDs 16276 and 16277, align with the two positive detections of J1325 with \textit{XMM-Newton}. These observations clearly show a short $\sim 8$ month outburst in 2014. The source flux rises over the course of the two \textit{XMM-Newton} observations in January and February 2014. The source flux may have peaked around the time of the first \textit{Chandra} detection in April 2014, but promptly fades after this, making this $\sim 8$ month outburst the only major outburst displayed by J1325. The third \textit{Chandra} detection of J1325 (ObsID 19747, May 2017) captured the source at a lower flux than those measured for the 2014 outburst. If the source was re-brightening or fading during this observation, that could explain the lower observed flux. However, the lack of corroborating data means we cannot confirm this. Another explanation is that this detection formed part of a mini-outburst similar to those observed in X-ray binaries, where mini-outbursts are often observed some time after a major outburst (see e.g. \citealt{MD2017} and \citealt{Ozbey2022}, respectively, for discussions of mini-outbursts in V404 Cygni and \textit{MAXI} J1820$+$070).

\vspace*{-0.3cm}
\subsubsection{{\em{Swift}}-XRT Monitoring}
On 2024 Feb 02, \textit{Swift} observed NGC 5128 for as part of a long-term monitoring program (ObsID 00049671009, TargetID 49671; see also Table \ref{tb:obs}), in which we detected an X-ray point source at the position of J1325 with an X-ray luminosity of $4.37 \times 10^{38}$ erg s$^{-1}$ (0.011 cts s$^{-1}$, converted to a flux using the same model as above (consistent with the best-fitting spectral model of the fainter \textit{XMM-Newton} detection; see Section \ref{sx:xmmspec}), indicating that J1325 was re-brightening (Figure \ref{fig:Long_Term_Lightcurve}, green diamond). To confirm whether the source was returning to outburst, we followed up with a 2.2 ks \textit{Swift} ToO on 2024-05-23 (ObsID 00015226011, TargetID 15226), under the premise that if the source was entering a major outburst, it should be detectable for $\sim 8$ months. However, J1325 was undetected. Subsequently, we checked the archival \textit{Swift-XRT} exposures taken within a few months of ObsID 00049671009 to place limits on the duration of the February 2024 rebrightening. There were four \textit{Swift-XRT} exposures prior to positive detection on 2024 Feb 02. These are ObsIDs 00015226009 (2.46 ks, 2023 Dec 31), 00015226010 (1.71ks, 2024 Jan 21), 00049671007 (1.56ks, 2024 Jan 28) and 00049671008 (0.39 ks, 2024 Feb 01). J1325 was undetected in all four exposures, although the non-detection on 2024 Feb 01 may be a result of the short exposure. Additionally, there was a \textit{Swift-XRT} exposure on 2024 Feb 12 (ObsID 00049671010, 0.065 ks), in which the source was also undetected. Therefore, J1325 increased in brightness on a few-day timescale between 2024-01-28 and 2024-02-01 and would have remained bright for less than $\sim 4$ months. This confirms that the source only briefly brightened, perhaps similar to the short rebrightening observed by \textit{Chandra} in 2017. Finally, we consult the \textit{Swift-XRT} X-ray Point Source Catalogue, 2SXPS \citep{Evans2020}, for any known source within $20$ arcseconds of J1325, but find no matches.

\begin{table*}
\begin{tabular}{c c c |  c c c c | c c c c }
\multicolumn{3}{| c |}{} & \multicolumn{4}{| c |}{Model 1: \texttt{tbabs(tbabs*powerlaw)}} & \multicolumn{4}{| c |}{Model 2: \texttt{tbabs(tbabs*diskbb)}}\\
\hline
ObsID & F$_{\rm{X}}$ & L$_{\rm{X}}$ & N$_{\rm{H}}$  & $\Gamma$ & $\chi$ & $\nu$ & N$_{\rm{H}}$  & T$_{\rm{in}}$ & $\chi$ & $\nu$\\
& [$10^{-13}$~erg cm$^{-2}$ s$^{-1}$] & [$10^{39}$~erg s$^{-1}$] & [$10^{22}$] & & & & [$10^{22}$] & [keV] & \\
\hline
\vspace*{+0.1cm}
0724060701 & 5.347 $\pm 0.073$ & 0.924 $\pm 0.013$ & 0.220 $\pm^{0.016}_{0.015}$ & 3.51 $\pm 0.09$ & 159.77 & 148 & $\leq 0.002$ & 0.431 $\pm^{0.008}_{0.007}$ & 219.15 & 148 \\
\vspace*{+0.1cm}
0724060801 & 9.840 $\pm^{0.130}_{0.131}$ & 1.701 $\pm^{0.023}_{0.022}$ & 0.146 $\pm 0.012$ & 2.69 $\pm 0.06$ & 170.71 & 169 & $\leq 0.001$ & 0.601 $\pm 0.013$ & 273.06 & 169 \\
\vspace*{+0.1cm}
16276 & 27.91 $\pm_{0.71}^{1.37}$ & 4.826 $\pm_{0.123}^{0.236}$ & 0.399 $\pm 0.07$ & 1.83 $\pm^{0.10}_{0.09}$ & 23.49 & 26 & 0.084 $\pm 0.005$ & 1.642 $\pm^{0.118}_{0.105}$ & 25.64 & 26 \\

\bottomrule
\end{tabular}
\caption{\label{tb:single} Best fitting spectral parameters and corresponding $1~\sigma$ errors obtained when fitting \textit{XMM-Newton} ObsIDs 0724060701 and 0724060801 and \textit{Chandra} ObsID 16276 in \textsc{xspec} with the spectral models \texttt{tbabs(tbabs*powerlaw)} and \texttt{tbabs(tbabs*diskbb)}. Here the first \texttt{tbabs} component is fixed to a value of $8.41 \times 10^{20}$ atom cm$^{-2}$, as determined by the Colden Galactic Neutral Hydrogen Density Calculator. The presented X-ray flux is the $0.5 - 10$ keV observed flux determined from the powerlaw spectral model using \texttt{flux} within \textsc{xspec}, and is converted to an observed luminosity assuming a distance of $3.8$ Mpc \citep{Harris2010}. For the \textit{XMM-Newton} observations, the spectra from the three EPIC detectors are fit simultaneously and the fluxes are combined EPIC values, determined by averaging the individual detectors fluxes.}
\end{table*}

\begin{table*}
\setlength{\tabcolsep}{4pt}
\begin{tabular}{c | c c c c c c | c c c c c c}
\multicolumn{1}{| c |}{} & \multicolumn{6}{| c |}{Model 3: \texttt{tbabs(tbabs(powerlaw + diskbb))}} & \multicolumn{6}{| c |}{Model 4: \texttt{tbabs(tbabs(diskbb + diskbb))}}\\
\toprule
ObsID & N$_{\rm{H}}$ & $\Gamma$ & T$_{\rm{in}}$ & Ratio & $\chi$ & $\nu$ & N$_{\rm{H}}$ & T$_{\rm{in}}$ & T$_{\rm{in}}$ & Ratio & $\chi$ & $\nu$\\
& [$10^{22}$] & & [keV] & & & & [$10^{22}$]& [keV] & [keV] & & & \\
\hline
\vspace*{+0.1cm}
0724060701 & 0.223 $\pm^{0.039}_{0.027}$ & 3.41 $\pm^{0.13}_{0.12}$ & 0.150 $\pm^{0.053}_{0.035}$ & 8.63 & 157.78 & 146 & 0.083 $\pm^{0.036}_{0.029}$ & 0.691 $\pm^{0.068}_{0.056}$ & 0.226 $\pm^{0.029}_{0.026}$ & 0.90 & 147.11 & 146 \\
\vspace*{+0.1cm}
0724060801 & 0.136 $\pm^{0.036}_{0.029}$ & 2.80 $\pm^{0.23}_{0.21}$ & 0.862 $\pm^{0.21}_{0.23}$ & 5.84 & 165.24 & 167 & 0.012 $\pm^{0.021}_{0.012}$ & 1.064 $\pm^{0.130}_{0.093}$ & 0.337 $\pm^{0.043}_{0.044}$ & 1.65 & 152.36 & 167 \\
\vspace*{+0.1cm}
16276 & 1.118 $\pm^{0.573}_{0.518}$ & 2.127 $\pm^{0.245}_{0.156}$ & 0.149 $\pm^{0.067}_{0.036}$ & 0.48 & 21.83 & 24 & 0.658 $\pm^{0.346}_{0.378}$ & 1.681 $\pm^{0.412}_{0.223}$ & 0.270 $\pm^{0.251}_{0.066}$ & 1.57 & 22.12 & 24 \\
\bottomrule
\end{tabular}
\caption{\label{tb:double}Best fitting spectral parameters and corresponding $1~\sigma$ errors obtained when fitting \textit{XMM-Newton} ObsIDs 0724060701 and 0724060801 and \textit{Chandra} ObsID 16276 in \textsc{xspec} with the spectral models \texttt{tbabs(tbabs(powerlaw + diskbb})) and \texttt{tbabs(tbabs(diskbb + diskbb})). Here the first \texttt{tbabs} component is fixed to a value of $8.41 \times 10^{20}$ atom cm$^{-2}$, as determined by the Colden Galactic Neutral Hydrogen Density Calculator. Here, we define the ratio as the $0.5 - 10$ keV flux of the first continuum model component divided by the  $0.5 - 10$ keV flux of the second continuum model component, where the fluxes of each component are computed with \texttt{cflux}.}
\end{table*}

\vspace*{-0.5cm}
\section{Two-Stage Search for X-ray Pulsations}
\label{sx:Pulse}

\begin{figure}
    \centering
    \includegraphics[width=\columnwidth]{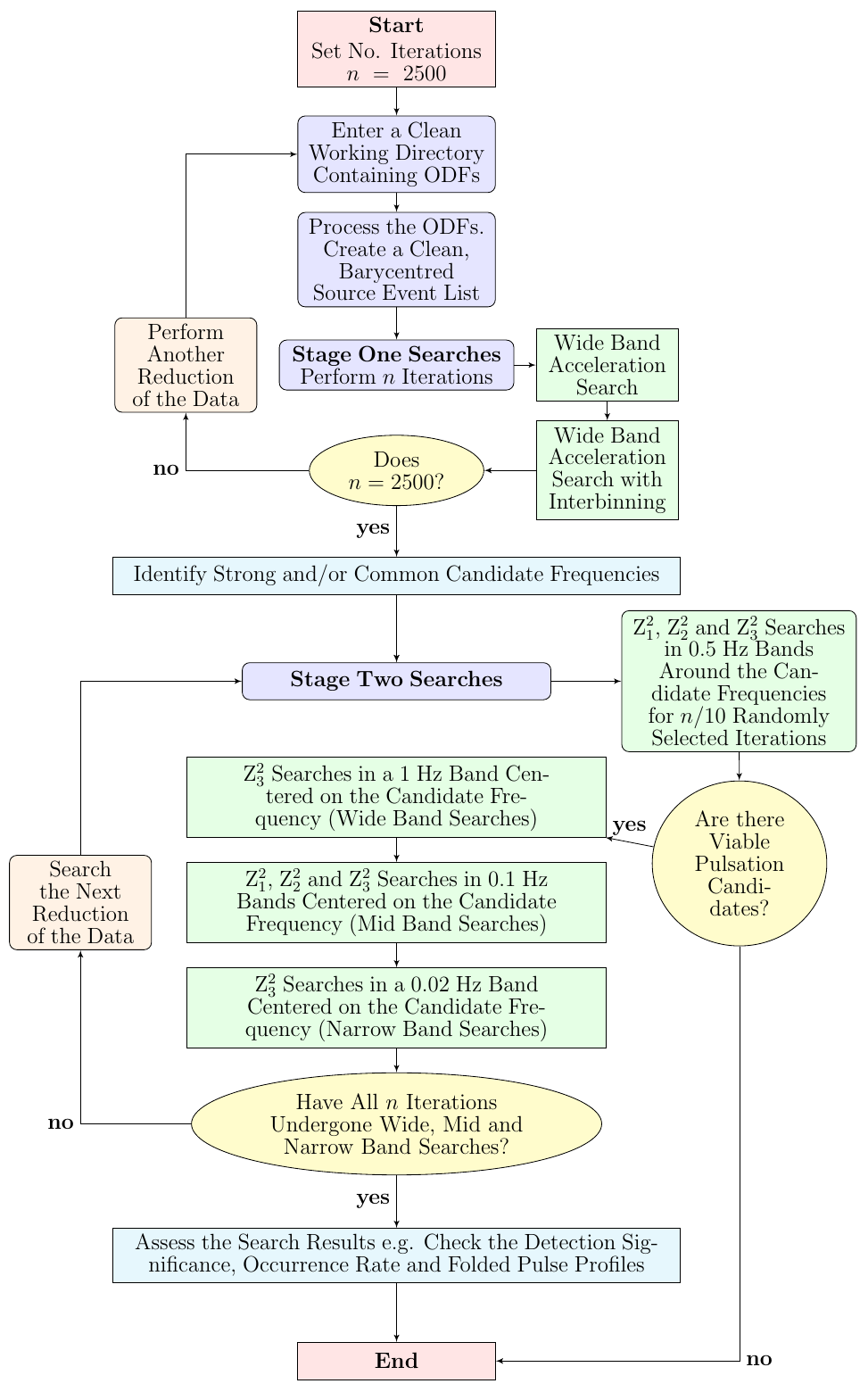}
    \vspace*{-0.25cm}
    \caption{A flow diagram showing the two stage pulsation search procedure implemented in this work.}
    \label{fig:Flow}
\end{figure}

The high spin-up rates typically exhibited by PULXs can make detecting pulsations challenging, especially if the source is distant or if the pulsations are weak or transient. To overcome these issues, we implement a two-stage search approach using \textsc{hendrics} version 8.1 \citep{Bachetti2018}, a set of command-line tools based on the \textsc{stingray} Python package (\citealt{Huppenkothen2019, Bachetti2024}) to perform temporal analysis of X-ray data. Our approach builds upon previous search strategies, performing both Fourier domain acceleration searches and Z$^{2}$ searches (see e.g. \citealt{Quintin2021}). However, it is novel in its considerations of the reproducibility of pulsation candidates. In this section, we describe the two search stages and our results in the context of reproducing and verifying our initial detection of X-ray pulsations at $1.27$ Hz in the $1 - 10$ keV band from J1325 in \textit{XMM-Newton} ObsID 0724060801 (see Section \ref{sx:Data}). We summarise the search results in Table \ref{tb:searches}.

\subsection{Motivation for Performing Multiple Iterations}
\label{sx:iterations}

\begin{figure}
    \centering
    \includegraphics[width=0.95\columnwidth]{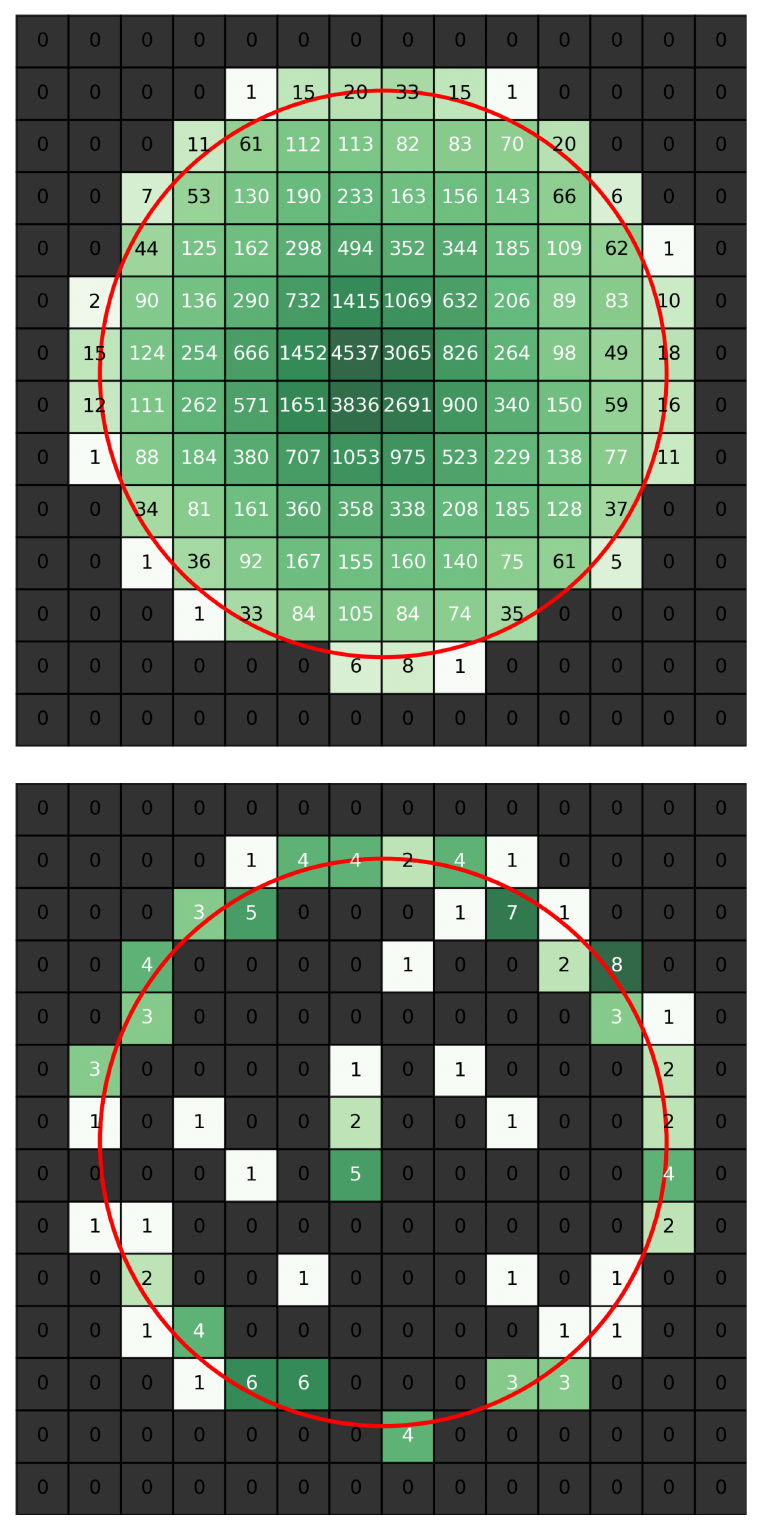}
    \vspace*{-0.25cm}
    \caption{A comparison of the counts per pixel within a selected source region when completing multiple reductions of \textit{XMM-Newton} data. Top: The number of counts per pixel for a single data reduction. Bottom: A comparison of two independent reductions, showing the number of unmatched counts (i.e., events present in one reduction that do not have a counterpart with similar arrival time and energy in the other). A band of discrepancies can be seen around the edge of the source region, where randomisation in each event's sub pixel position during reduction causes different events to be selected each time.}
    \label{fig:reduction_comparison}
\end{figure}

As shown in Figure \ref{fig:Flow}, the start of the search procedure involves defining the value, $n$, which is the number of times each set of ODFs is reduced (also referred to as iterations), through to the point of creating a clean source event list (see Section \ref{sx:Data}). Each data reduction then undergoes stage one and stage two searches. While re-reducing the data $n$ times may appear as an unnecessary and time-consuming step in our search procedure, we will demonstrate throughout this section that it is critical when determining the true strength of any marginal pulsation signals identified in \textit{XMM-Newton} observations. The reason for this is that when initialising \textsc{xmm-sas} to reduce \textit{XMM-Newton} data, as we have done in Section \ref{sx:Data}, a random seed variable, \texttt{SAS$\_$RAND$\_$SEED}, is set, which initialises the single global random number generator used in many \textsc{xmm-sas} reduction tasks. While the user can provide an integer value for the random seed variable, this is not common practice. If not specified, a pseudo-random number, based on the local clock time, is assigned to the random seed variable. As a result, processing the same dataset twice, even with identical command line arguments, will not provide the exactly same results \footnote{\url{https://xmm-tools.cosmos.esa.int/external/sas/current/doc/taskmain/node3.html}}. Within \textsc{xmm-sas}, the randomisation is introduced to avoid event properties being highly quantised. However, this means that each time a set of ODFs is processed, the resulting event lists differ. An example of these differences is shown in Figure \ref{fig:reduction_comparison}. Here, we see a band of event discrepancies around the edge of the selected circular source region, as well as some variations within the region. These variations arise from the randomisation of event sub-pixel positions introduced by \texttt{epproc}, which can cause individual events to shift inside or outside the extraction region between reductions, due to the region passing through the pixel or between neighbouring pixels. These shifts depend on the integer value of \texttt{SAS$\_$RAND$\_$SEED}. We initially uncovered these reproducibility issues when trying (and failing) to exactly replicate the initial $1.27$ Hz detection noted in Section \ref{sx:Data}. To reproduce an exact set of results, \texttt{SAS$\_$RAND$\_$SEED} must be defined. However, this raises the somewhat unanswerable question of what value to assign to the random seed variable. To understand the impact of this variable on X-ray pulsation searches with \textsc{hendrics} and to determine the true strength of our marginal $1.27$ Hz signal, we choose $n = 2500$ to compile a distribution of pulsation frequencies and strengths from which to draw our conclusions.

\subsection{Stage One: Fourier Domain Acceleration Searches}
\label{sx:Accel}

\begin{figure*}
    \centering
    \includegraphics[width=\textwidth]{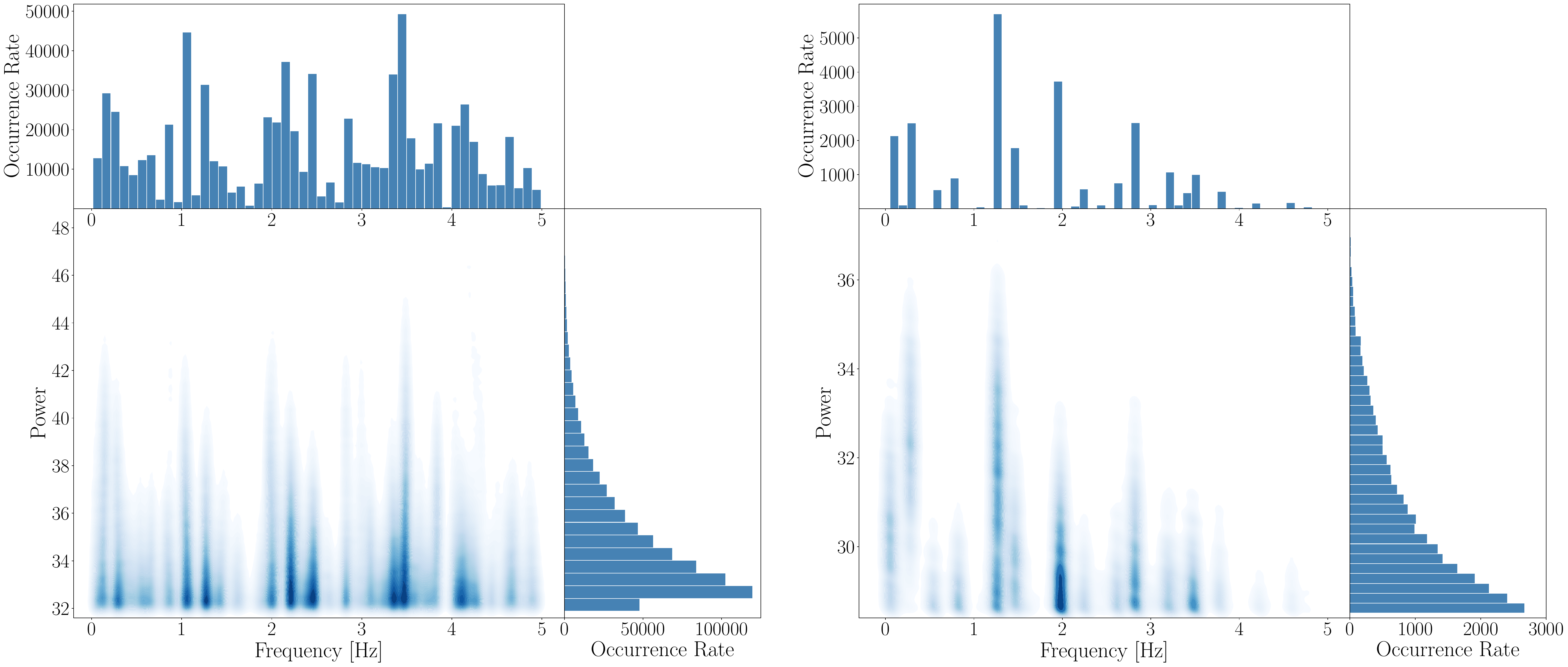}
    \vspace*{-0.5cm}
    \caption{Histograms showing the candidate pulsation frequencies and corresponding powers returned when conducting 2500 data reduction iterations and acceleration searches of \textit{XMM-Newton} ObsID 0724060801, in the frequency range $0.01 - 5$ Hz, with (left) and without interbinning (right). The acceleration search with interbinning returns roughly 350 candidates per data reduction iteration, many of which are either duplicates or false positives. Therefore, we present only the top $300$ independent candidates in the left-hand panel. The acceleration search without interbinning returns approximately $10 - 15$ candidates per data reduction iteration so we plot all candidates in the right-hand panel. Here, the occurrence rate represents the total number of times a frequency appears within each $0.1$~Hz frequency bin (top) or within each $0.324$ width power bin (side) across the 2500 iterations.}
    \label{fig:Accelsearch}
\end{figure*}

We conduct our stage one, Fourier domain acceleration searches within \textsc{hendrics} using the \texttt{accelsearch} routine, which takes a single Fast Fourier Transform (FFT) and convolves it with responses corresponding to different values of acceleration (spin up / spin down; see also \citealt{Ransom2002b}). The benefit of performing an acceleration search of any type is speed; this technique enables fast, untargeted searches for pulsations, which is particularly beneficial when searching for PULXs, as the neutron star rotation rate is initially unknown. It is for this reason that we conduct the acceleration searches first. For each data reduction iteration, we perform two acceleration searches. As in Figure \ref{fig:Flow}, the first is a standard acceleration search and the second is an acceleration search with interbinning, both of which we conduct over a wide frequency band of $0.01 - 5$ Hz with $\dot{f}$ kept at the default value of $1$. In some cases, a wider frequency band, e.g. $0.01 - 7$ Hz, may be appropriate (e.g. \citealt{Quintin2021}) and an upper limit of $6.8$ Hz is possible with Nyquist sampling given the $0.073$ ms readout time of EPIC-PN. However, we select an upper limit of $5$ Hz to reduce both the computational time of each acceleration search and the noise associated with the readout time of EPIC-PN, the former of which is an important consideration when conducting thousands of iterations. 

Each of the acceleration searches has its benefits. The standard acceleration search is most effective when the observation duration is much less than the orbital period \citep{Ransom2002b}. We perform this search first because it will identify any strong signals that are present. If the standard acceleration search returns a strong/dominant pulsation candidate, one could reasonably bypass the subsequent acceleration search with interbinning; however, this decision should occur on a case-by-case basis. For example, given the standard acceleration results presented in the right hand panel of Figure \ref{fig:Accelsearch}, which shows two dominant candidates at $1.27$ and $\sim2$ Hz respectively, one could proceed to the stage two searches without conducting an acceleration search with interbinning.

Interbinning is a computationally inexpensive form of Fourier interpolation where the signal in the two closest integer frequency bins approximates the Fourier transform response at half-integer frequencies (\citealt{vanderKlis1989}, see also \citealt{Ransom2002b}). This technique overcomes the loss of sensitivity in the power density spectrum near the edges of the spectral bins, reducing the scalloping effect. Consequently, an acceleration search with interbinning is more effective at locating weak or transient signals than a standard acceleration search, particularly for signals that lie close to the edges of the integer bins. For this reason, our default search procedure includes both a standard acceleration search and an interbinned acceleration search, as the latter can locate any marginal candidates that the standard search may miss. However, as discussed by \citet{Ransom2002b}, the interbins possess different properties from the integer FFT bins. These include different noise properties, which impact the calculation of the significance of interbin powers. Also, the interbins are not independent Fourier trials, as they are correlated with the integer Fourier bins that created them. These contribute to the downsides of an interbinned acceleration search, including a high false positive rate that arises from modifications to the white noise level and the clustering of candidates around a genuine signal. Therefore, while verifying the results of any acceleration search with a focused Z$^{2}$ search is important, it becomes essential for any acceleration search with interbinning. For these reasons, excluding the acceleration search with interbinning from the search procedure is recommended for any case where the standard acceleration search returns viable pulsation candidates.

The stage one search results are summarised in Figure \ref{fig:Accelsearch}, where we see stark differences between the results with interbinning (left-hand panel) and without (right-hand panel). As expected, the interbinned acceleration search returns many more candidates than the standard acceleration search, with candidate frequencies identified across the entire search range. Roughly $350$ candidates get returned from each iteration of the acceleration search with interbinning, versus approximately $15$ candidates from each standard acceleration search iteration. Many candidates found by the interbinned acceleration search are likely false positives, and many of the candidates returned are duplicates, so we only plot the strongest $300$ independent candidates from each interbinned search iteration in Figure \ref{fig:Accelsearch}. For the standard search, we plot all candidates from each iteration. The acceleration search with interbinning does not return a single stand-out signal. Instead, there are noticeable hot spots in the frequency bins centred on $4.14$, $3.64$, $3.54$, $1.26$ and $1.06$ Hz. There is also a broad cluster between $2.0$ and $2.5$ Hz. The signal strengths (powers) are skewed towards the lower end of the range, with few candidates surpassing powers of $\sim 40$, although we note that the signals in the bin centred on $3.64$ Hz surpass a power of $40$ most often. It is also the signal with the highest occurrence rate (i.e. the signal identified most often), closely followed by the signal at $\sim 1.1$ Hz. The standard acceleration search, in comparison, finds far fewer signals and, as shown in the right-hand panel of Figure \ref{fig:Accelsearch}, there are two stand-out frequencies at $1.27$ and $1.99$ Hz. The former has the highest occurrence rate and often appears with high powers, while the latter is a weaker and broader detection. Both of these signals are present in the interbinned acceleration search, falling into the $\sim 1.26$ Hz and $\sim 2.0$ Hz hot spots, respectively (see also Table \ref{tb:searches}). Interestingly, the standard acceleration search does not find a peak at $\sim 4.0$ Hz, and the signal at $\sim 3.5$ Hz does not appear significant. 

To determine which frequencies to use in the stage two searches, we assess both acceleration searches against three factors: the occurrence rate, the power distribution and the overlaps between the two acceleration searches. From the interbinned search, we discard the $4.14$ Hz bin because it has the lowest occurrence rate and does not appear in the standard acceleration search. Neither the $3.54$ nor $3.64$ Hz peaks appear in the standard search results. However, the latter bin has the highest occurrence rate from the interbinned search, so we chose to keep it for the next stage. The bin centred on $1.26$ Hz contains the best pulsation candidate as it aligns with the $1.27$ Hz bin from the standard search, which is also the bin with the highest signal strength and occurrence rate. We note that this is consistent with our initial finding (see Section \ref{sx:Data}). We also keep the broad detections around $\sim 2$ Hz because they appear in both searches. Therefore, we will conduct our preliminary stage-two searches in frequency bands centred on $1.27$, $2.0$, $2.3$ and $3.5$ Hz.   

Additionally, these results demonstrate a clear benefit of our repetitive search procedure -- running multiple data reduction iterations allows us to overcome the intrinsic randomisation in \textsc{xmm-sas} and select the best frequencies to proceed to stage two. For example, the strongest signal from the interbinned acceleration search of data reduction iteration 732 was $4.171$ Hz at a power of $45.6$. However, it does not have a high occurrence rate and does not often appear with a high power. Therefore, it is likely a false positive candidate. Building up a distribution of signals allows us to remove obvious false positive signals and focus the stage two searches on the more promising candidates. 

\subsection{Stage Two: Z$^2$-Searches}
In stage two of our procedure, we conduct several targeted Z$^{2}$ searches to ascertain whether any of the frequencies carried forward from stage one are viable pulsation candidates and subsequently assess whether any of the pulsation candidates deemed viable are likely to be pulsation signals with astrophysical origin. Throughout this and subsequent sections, we will refer to three types of search results: a candidate, a top candidate and a confirmed candidate. Here, a `candidate' refers to any frequency returned by the search. A `top candidate' refers to the strongest (highest power) frequency returned by the search. This will appear as the first candidate in the lists returned by \textsc{hendrics}. Lastly a `confirmed' candidate is a frequency returned by the search that surpasses the $99.9~\%$ statistical detection limit set by \textsc{hendrics}, indicating that pulsations were detected at the candidate frequency.

As in Figure \ref{fig:Flow}, we first run $n / 10 $ preliminary $Z^{2}_{1}$, $Z^{2}_{2}$ and $Z^{2}_{3}$ searches in $0.5$ Hz bands centred on frequencies carried forward from the stage one searches to determine whether any of these are viable pulsations candidates. The iterations searched here are selected randomly from the total number of data reductions already performed. Here, a viable pulsation candidate is a frequency that has a high occurrence rate and exhibits a sinusoidal folded pulse profile (consistent with the pulse profiles of known PULXs). A viable signal may also surpass the $99.9$ per cent statistical detection threshold within \textsc{hendrics}, but we do not exclude signals below this threshold. Ultimately, we must use discretion at this stage, and not all three criteria need to be satisfied to pursue a frequency further.

Should a viable pulsation candidate emerge, we continue through the stage two searches (see Figure \ref{fig:Flow}) and employ a set of targeted Z$^{2}$ searches to explore how dominant our candidate frequency is and to investigate how the candidate varies when varying the search parameters, specifically the frequency range and the number of harmonics used. When conducted over a narrow frequency band, Z$^{2}$ searches will return results with higher strengths simply because the search doesn't know about the other n-trials. In other words, the search algorithm is only aware of the frequencies within its specific search range (where the n-trials are the number of independent frequencies in that frequency range). Therefore, using a narrow band can artificially increase the strength of a pulsation signal and may lead to false positive detections. To avoid this, we conduct Z$^{2}$ searches in $1$ (wide), $0.1$ (mid), and $0.02$ (narrow) Hz bands centred on the candidate frequencies. If a signal is present in all three bands, especially if it appears in the wide-band searches, it is more likely to be a genuine signal. Similarly, varying the number of harmonics in the searches assesses whether the candidate frequency is a bona fide pulsation as the signals returned from a $Z^{2}$ search are classified based on their harmonic content. Therefore, if a Z$^{2}_{3}$ search locates a signal with three harmonics, it will be returned with a higher power than a signal from the search found with only two harmonics. As a result, performing a Rayleigh test (Z$^{2}_{1}$ searches) typically returns a more varied list of signals, generally with lower powers, and one should confirm the signals with higher harmonic searches. On the other hand, a Z$^{2}_{3}$ search will naturally hone in on a candidate with the highest harmonic content, which could be a genuine pulsation signal or an instrumental signal (e.g. the readout signal), and easily distinguished by the shape of the folded pulse profile.

\subsubsection{Preliminary Z$^2$-Searches}
\begin{figure*}
    \centering
    \includegraphics[width=\textwidth]{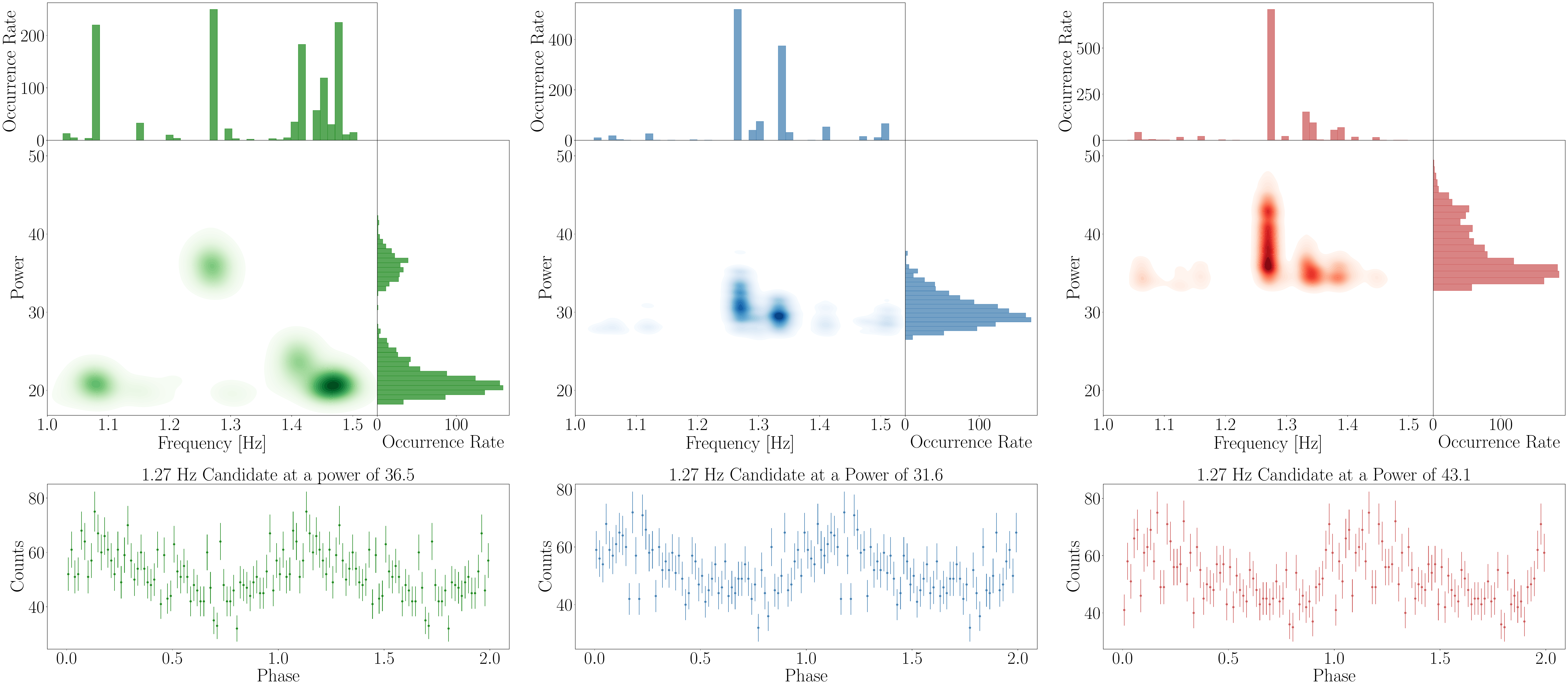}
    \vspace*{-0.3cm}
    \caption{Histograms showing the candidate pulsation frequencies and corresponding powers returned when conducting $Z^{2}_{1}$ (left), $Z^{2}_{2}$ (middle) and $Z^{3}_{2}$ searches of 250 data reduction iterations of \textit{XMM-Newton} ObsID 0724060801 randomly selected from the 2500 data reduction iterations performed during stage one. The searches are conducted in the frequency range $1.02 - 1.52$ Hz. The lower subplot of each panel shows the folded pulse profile of the strongest candidate returned from the twentieth iteration. The frequency and power of the candidate is given above each subplot. Here, the occurrence rate is the total number of times a candidate appears within each frequency (top) or power (side) bin across the 250 iterations.}
    \label{fig:Prelim_Band1}
\end{figure*}

\begin{figure*}
    \centering
    \includegraphics[width=\textwidth]{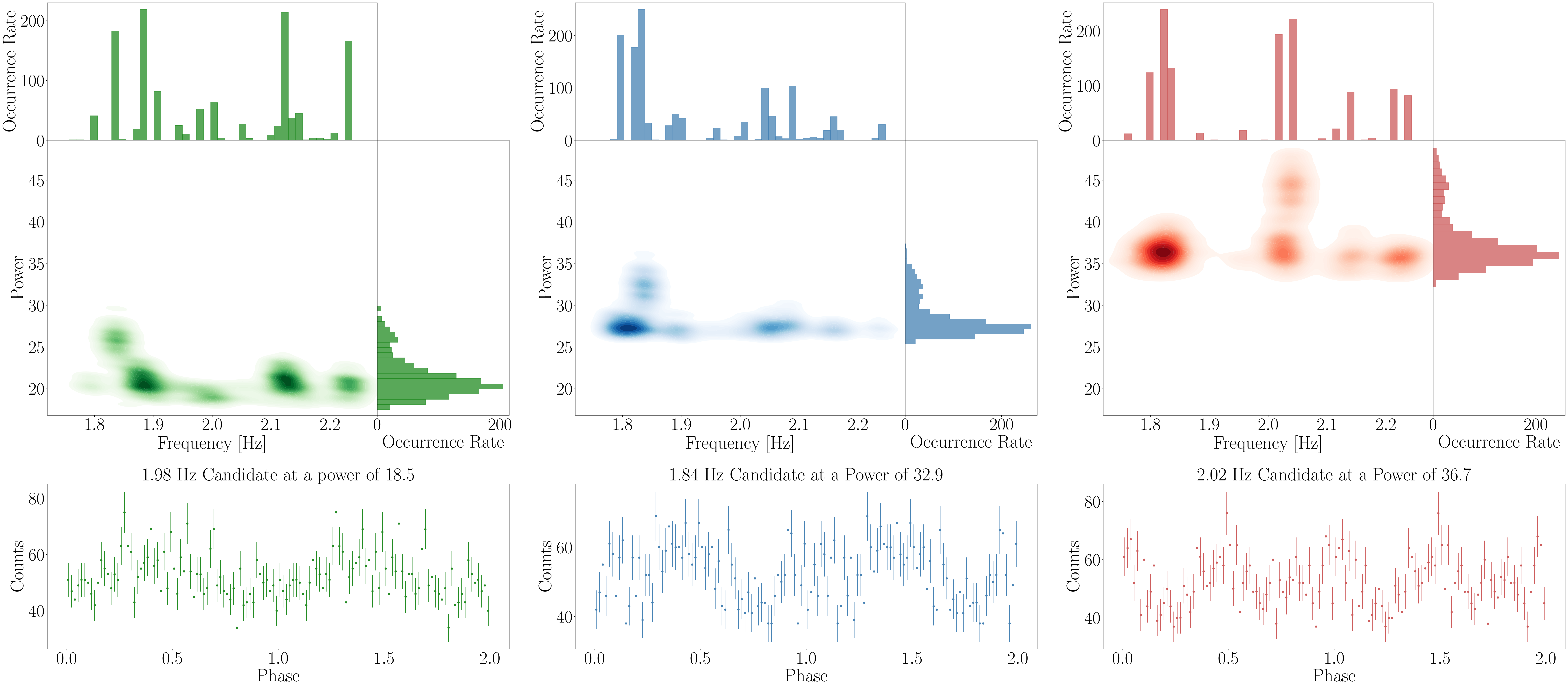}
    \vspace*{-0.3cm}
    \caption{Histograms showing the candidate pulsation frequencies and corresponding powers returned when conducting $Z^{2}_{1}$ (left), $Z^{2}_{2}$ (middle) and $Z^{3}_{2}$ searches of 250 data reduction iterations of \textit{XMM-Newton} ObsID 0724060801 randomly selected from the 2500 data reduction iterations performed during stage one. The searches are conducted in the frequency range $1.75 - 2.25$ Hz. The lower subplot of each panel shows the folded pulse profile of the strongest candidate returned from the twentieth iteration. The frequency and power of the candidate is given above each subplot. Here, the occurrence rate is the total number of times a candidate appears within each frequency (top) or power (side) bin across the 250 iterations.}
    \label{fig:Prelim_Band2}
\end{figure*}

\begin{figure*}
    \centering
    \includegraphics[width=\textwidth]{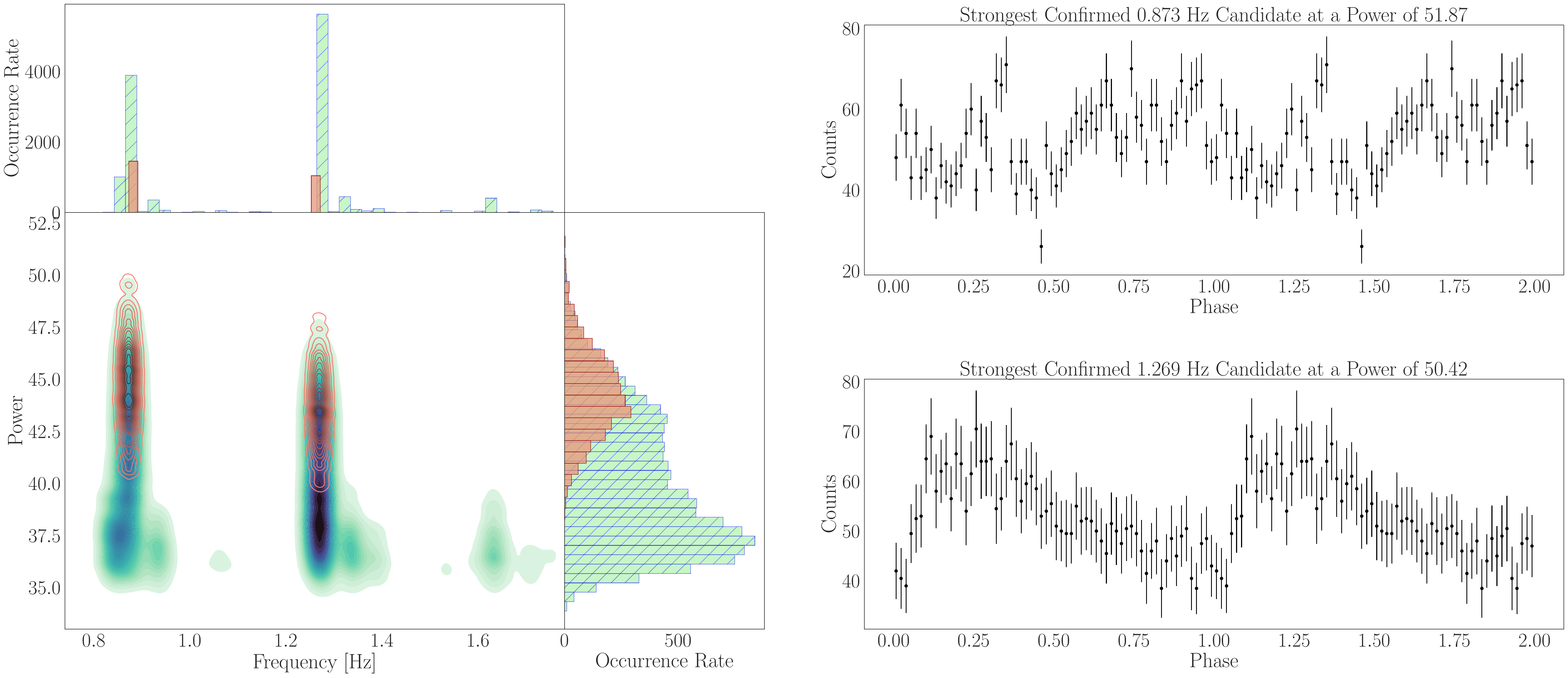}
    \vspace*{-0.35cm}
    \caption{Left: Histograms showing the candidate pulsation frequencies and corresponding powers returned when conducting 2500 data reduction iterations and performing $Z^{2}_{3}$ searches of \textit{XMM-Newton} ObsID 0724060801, in the frequency range $0.77 - 1.77$ Hz. Here the hatched histograms show the frequency and power distributions of all returned candidates (approximately 15 candidates per search iteration) and the orange histograms show the same distributions for just the top (strongest) candidate from all search iterations. Right: The folded pulse profile of the strongest $0.873$ Hz candidate (top) and the strongest $1.269$ Hz candidate (bottom), which are respectively from iterations 1728 and 150. Here, the occurrence rate is the total number of times a candidate appears within each frequency (top) or power (side) bin across the 2500 iterations.}
    \label{fig:Wide}
\end{figure*}

We randomly select $n / 10$ (250) data reduction iterations from those performed during stage one and and conduct $Z^{2}_{1}$, $Z^{2}_{2}$ and $Z^{2}_{3}$ searches of each data reduction iteration in the following four frequency bands: $1.02 - 1.52$, $1.75 - 2.25$, $2.15 - 2.55$ and $3.25 - 3.75$ Hz. For conciseness, we show the results from the first two bands in Figures \ref{fig:Prelim_Band1} and \ref{fig:Prelim_Band2} respectively. Figures showing the results of searches in the latter two bands are provided as online supporting material. We also note that the reduced number of iterations is sufficient to obtain an output distribution representative of a search with $n$ iterations, but lowers the computational time required, which is beneficial when assessing several candidates. 

As shown in Figure \ref{fig:Prelim_Band1}, these preliminary searches identify the $1.27$ Hz candidates in all three cases. When looking at the $Z^{2}_{1}$ results (left), we see that the $1.27$ Hz candidate appears with similar occurrence rates to candidates at $1.08$ and $1.46$ Hz. However, the $1.27$ Hz candidate is the only one returned with powers over 30, and all of the $1.27$ Hz candidates returned by the $Z^{2}_{1}$ search surpass the $99.9~\%$ detection threshold, making them statistically confirmed pulsations within \textsc{hendrics} (see column 5 in Table~\ref{tb:searches}). In contrast, the $Z^{2}_{2}$ (middle) and $Z^{2}_{3}$ (right) searches, respectively, return zero and five statistically confirmed pulsation candidates at $1.27$ Hz, but the $1.27$ Hz signal is the dominant candidate in both cases. The lower panels of Figure \ref{fig:Prelim_Band1} show the folded pulse profiles of the top candidate from the twentieth data reduction iteration for each of the three searches. As shown in the figure, the top candidate is $1.27$ Hz in all three searches, and the pulse profiles are approximately sinusoidal in each case. When considering these folded pulse profiles along with the fact that the $1.27$ Hz signal appears in all three cases, it strongly suggests that $1.27$ Hz pulsations are present in this data set. Therefore, we will explore the $1.27$ Hz candidate in the detailed stage two searches.

In contrast, Figure \ref{fig:Prelim_Band2}, which shows the search results in the $1.75 - 2.25$ Hz band, shows that the candidate with the highest occurrence rate changes with the number of harmonics used for the search. For the $Z^{2}_{1}$, $Z^{2}_{2}$ and $Z^{2}_{3}$ these are $1.87$, $1.84$ and $1.82$ Hz respectively. While these frequencies are similar and potentially indicative of pulsations, we disfavour these candidates for a few reasons. Firstly, the candidates with the highest occurrence rates are often not the top candidates. This is demonstrated by the lower panels, which show the folded pulsed profile of the top candidate from the twentieth data reduction iteration for each of the three $Z^{2}$ searches. Furthermore, the top candidates returned by each of the three searches differ with each iteration. Figure \ref{fig:Prelim_Band2} also demonstrates that no one frequency is dominant, as several frequency bins have similar occurrence rates, which was not the case for the $Z^{2}_{2}$ and $Z^{2}_{3}$ searches shown in Figure \ref{fig:Prelim_Band1}. This could mean that the pulsation candidates in the $1.75 - 2.25$ Hz frequency range are weaker and thus more challenging to discern from the surrounding frequencies. However, it could also be that the signals are false positives. For these reasons, and because the folded pulse profiles are not clear sinusoids, we have sufficient evidence to exclude the candidates in the $1.75 - 2.25$ Hz frequency range from further searches. Similar arguments to those used here also allow us to exclude candidates in the $2.15 - 2.55$ and $3.25 - 3.75$ Hz from further stage two searches. Further details are provided in Table \ref{tb:searches} and in the supplementary material.

\subsubsection{Wide Band Search}
Figure \ref{fig:Wide} shows the results of our wide band ($0.77 - 1.77$ Hz) Z$^{2}_{3}$ search. Note that because a search with a higher harmonic content naturally returns the candidates with the highest number of harmonics, we choose to run only the Z$^{2}_{3}$ search in the wide frequency band, thereby reducing the number of candidates returned. In the left-hand panel of Figure \ref{fig:Wide}, the hatched (blue) data show all returned candidates (five candidates per data reduction iteration) while the orange data show only the strongest candidate from each iteration. As shown, there are two dominant frequencies -- $1.27$ and $0.87$ Hz -- the former of which is a candidate that we identified in our initial analysis (see Section \ref{sx:Data}), both the stage one searches (Figure \ref{fig:Accelsearch}) and the preliminary Z$^{2}$ searches (Figure \ref{fig:Prelim_Band1}). Among all returned candidates, $1.27$ Hz is the signal with the highest occurrence rate at $45.5~\%$, while the $0.87$ Hz signal has an occurrence rate of $30.8~\%$ (see also Table \ref{tb:searches}). Also apparent from this plot is the wide range of strengths returned for a single frequency; the 1.27 Hz signal is found with powers roughly in the range $35 - 48$, depending on which data reduction iteration we are searching, and thus the value of \texttt{SAS$\_$RAND$\_$SEED}. This demonstrates our reasoning for performing multiple data reductions (see Section \ref{sx:iterations}) as the interpretation of a candidate would naturally differ when returned with a power of $35$ compared to a power of $46$. However, the $0.87$ Hz signal is intriguing as it appears as the top candidate more often than the $1.27$ Hz signal and displays a similar distribution of powers (again, all attributed to a single frequency). Also, both acceleration searches found corresponding signals at $\sim 0.9$ Hz, albeit with low occurrence rates. However, as shown in the right-hand panels of Figure \ref{fig:Wide}, the $1.27$ Hz signal (bottom) displays a clear skewed sinusoidal folded pulse profile, consistent with a signal of astrophysical origin, while the 0.87 Hz signal does not. We also confirm that the signals are not resonances of one another. Therefore, we can confidently discard the $0.87$ Hz candidate and have increased confidence in the $1.27$ Hz candidate. 

\begin{figure}
    \centering
    \includegraphics[width=\columnwidth]{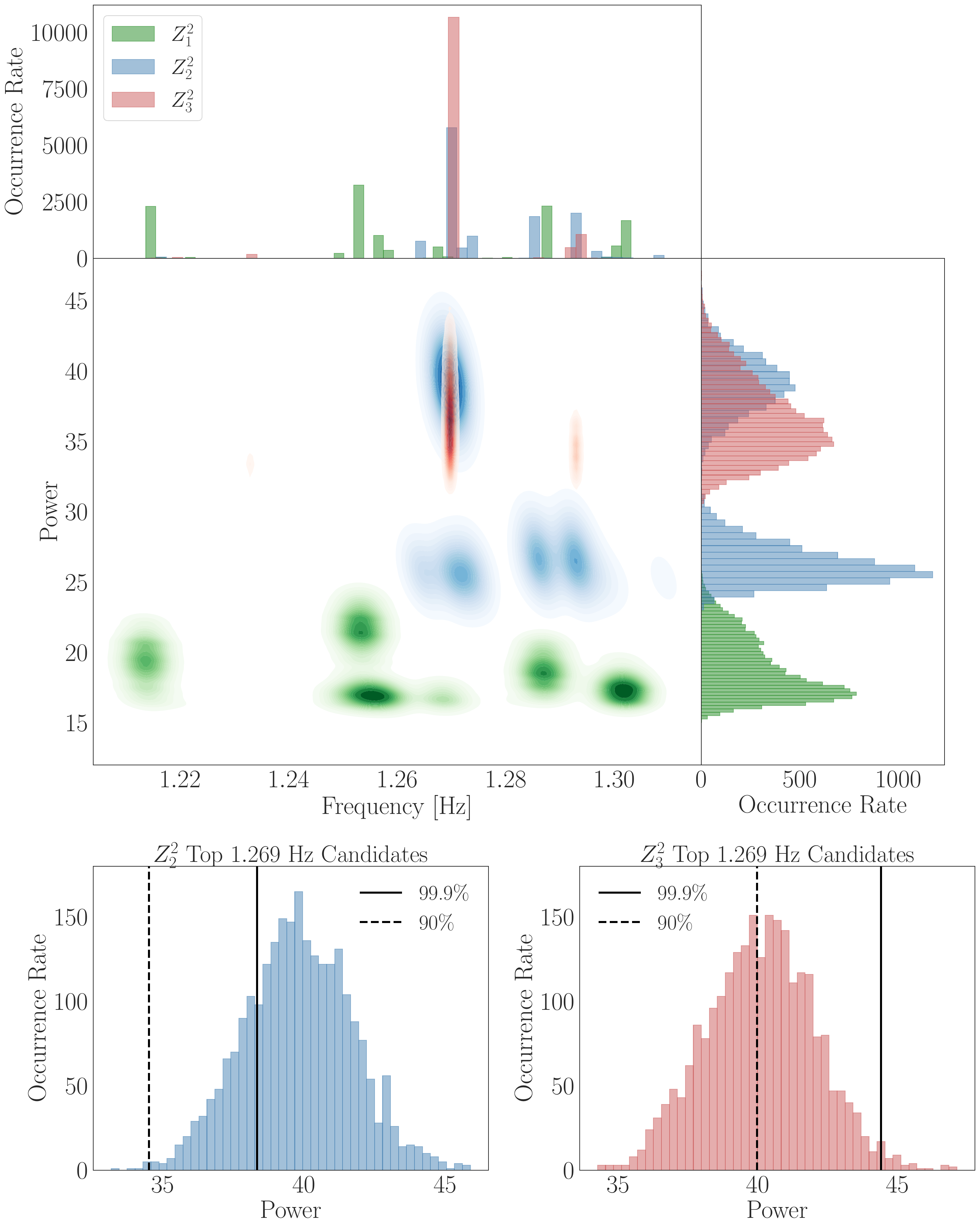}
    \vspace*{-0.3cm}
    \caption{Top: Histograms showing the candidate pulsation frequencies and corresponding powers returned from all candidates when conducting 2500 data reduction iterations and performing $Z^{2}_{1}$ (green),  $Z^{2}_{2}$ (blue) and $Z^{2}_{3}$ (red) searches of \textit{XMM-Newton} ObsID 0724060801, in the frequency range $1.21 - 1.31$ Hz. Bottom: Power distributions of the top candidates from the $Z^{2}_{2}$ (left) and $Z^{2}_{3}$ (right). The top candidate is always $1.269$ Hz for both the $Z^{2}_{2}$ and $Z^{2}_{3}$ searches. In both panels, the dashed and solid lines show the 90 per cent and 99.9 per cent statistical detection thresholds, respectively. For all 2D-histograms, the occurrence rate is the total number of times a candidate appears within each frequency or power bin across the 2500 iterations.}
    \label{fig:ZMid}
\end{figure}

\begin{figure}
    \centering
    \includegraphics[width=0.95\columnwidth]{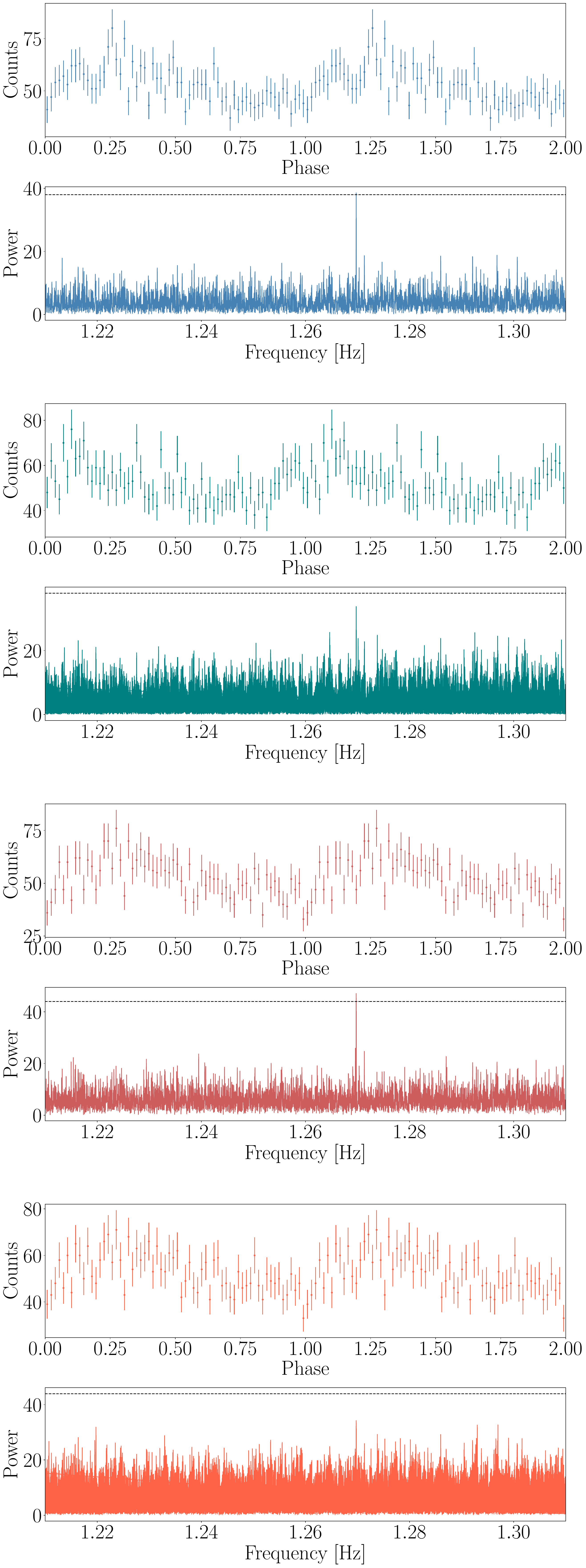}
    \vspace*{-0.15cm}
    \caption{Folded pulse profiles and frequency-power spectra of the strongest and weakest candidates returned by the $Z^{2}_{2}$ (rows one and two; blue/green),  $Z^{2}_{3}$ (rows three and four; red/orange) searches of \textit{XMM-Newton} ObsID 0724060801, in the frequency range $1.21 - 1.31$ Hz. The top candidate is always $1.269$ Hz and the dashed horizontal lines show the 99.9 per cent statistical detection threshold.}
    \label{fig:FreqSpec}
\end{figure}

\subsubsection{Mid-Band Search}
\label{sx:mid}
The results of our mid-band ($1.21 - 1.31$ Hz) searches are shown in Figure \ref{fig:ZMid} (see also Table \ref{tb:searches}), where we see the distribution of all returned candidates from the Z$^{2}_{1}$ (green), Z$^{2}_{2}$ (blue) and Z$^{2}_{3}$ (red) searches, respectively. The $Z^{2}_{1}$ searches return several signals across the frequency range, all of which appear with lower powers compared to the $Z^{2}_{2}$ and $Z^{2}_{3}$ searches. Four of these signals appear with similar occurrence rates, suggesting that the $Z^{2}_{1}$ search is not as effective in distinguishing between signals when the search band is narrowed and is more reliable when conducted over wider search ranges. Comparing Figures \ref{fig:Prelim_Band1} (left) and \ref{fig:ZMid}, which show the results from $Z^{2}_{1}$ searches centred on the $1.27$ Hz candidate frequency, in 0.5 and 0.1 Hz bands, respectively, demonstrates this. The former can locate $1.27$ Hz sinusoidal signals at $\geq 99.9 \%$ statistical significance over $250$ iterations. However, the latter does not statistically confirm the presence of pulsations at any frequency despite increasing the number of iterations by a factor of ten. These differences may arise if narrowing the frequency range of the search also narrows the search in $\dot{f}$, such that the overall parameter space search differs. However, we do not find any evidence to support this hypothesis. Thus, it is unclear what the exact cause of these differences is. Notably, the dominant signal returned from the mid-band $Z^{2}_{1}$ search is $1.253$ Hz, although the $1.27$ Hz signal is still found with powers $<20$. Therefore, we are cautious when narrowing the frequency band for a Rayleigh test because the results appear dependent upon the search parameters. Rayleigh test findings should, therefore, be verified through additional searches. 

In comparison, the $Z^{2}_{2}$ and $Z^{2}_{3}$ searches both find a $1.269$ Hz signal with higher powers. As is evident from Figure \ref{fig:ZMid} (middle and top, respectively), as the number of harmonics used in the search increases, the range of the frequencies returned decreases, allowing us to hone in on the dominant signal in the data. However, the significance of the $1.269$ Hz signal remains challenging to constrain. As seen in the top right histogram and both lower panels of Figure \ref{fig:ZMid}, the distribution of signal powers among all returned candidates is incredibly varied, particularly for the $Z^{2}_{2}$ searches (blue, Figure \ref{fig:ZMid}), where we see a two-peaked power distribution for the $1.269$ Hz signal. These results reiterate the need to search multiple data reduction iterations when assessing the significance of a marginal pulsation candidate. For example, the consecutive data reduction iterations $1365$ and $1366$ have timestamps separating them by just four minutes \footnote{The consecutive data reduction iterations are selected to emphasis how different the results can be depending on the value of the random seed variable. However, we note that the temporal separation between reductions has no impact on the randomisation beyond determining a different initial seed. Different seeds produce effectively uncorrelated reconstructions regardless of whether the runs are minutes or years apart.}. Recall also that the timestamp will define the random seed variable employed by \textsc{xmm-sas}. While iteration $1365$ returns a top candidate at $1.269$ Hz at a power of $34.56$, barely reaching the $90$ per cent statistical detection threshold, iteration $1366$ returned a top candidate at exactly the same frequency but at a significantly different power of $44.29$. While the first result could be easy to overlook on account of its lower power, the second result, which we note also surpasses the $99.9$ per cent statistical detection threshold in \textsc{hendrics}, appears more promising. However, performing only a single data reduction (which could be considered standard practice) would not show where in the power distribution the returned candidate falls. Therefore, trusting a single data reduction iteration is inadvisable as it can lead to incorrect conclusions about the presence of pulsations in marginal cases, and impact our understanding of the number of ULXs believed to host NSs. 

\begin{figure}
    \centering
    \includegraphics[width=\columnwidth]{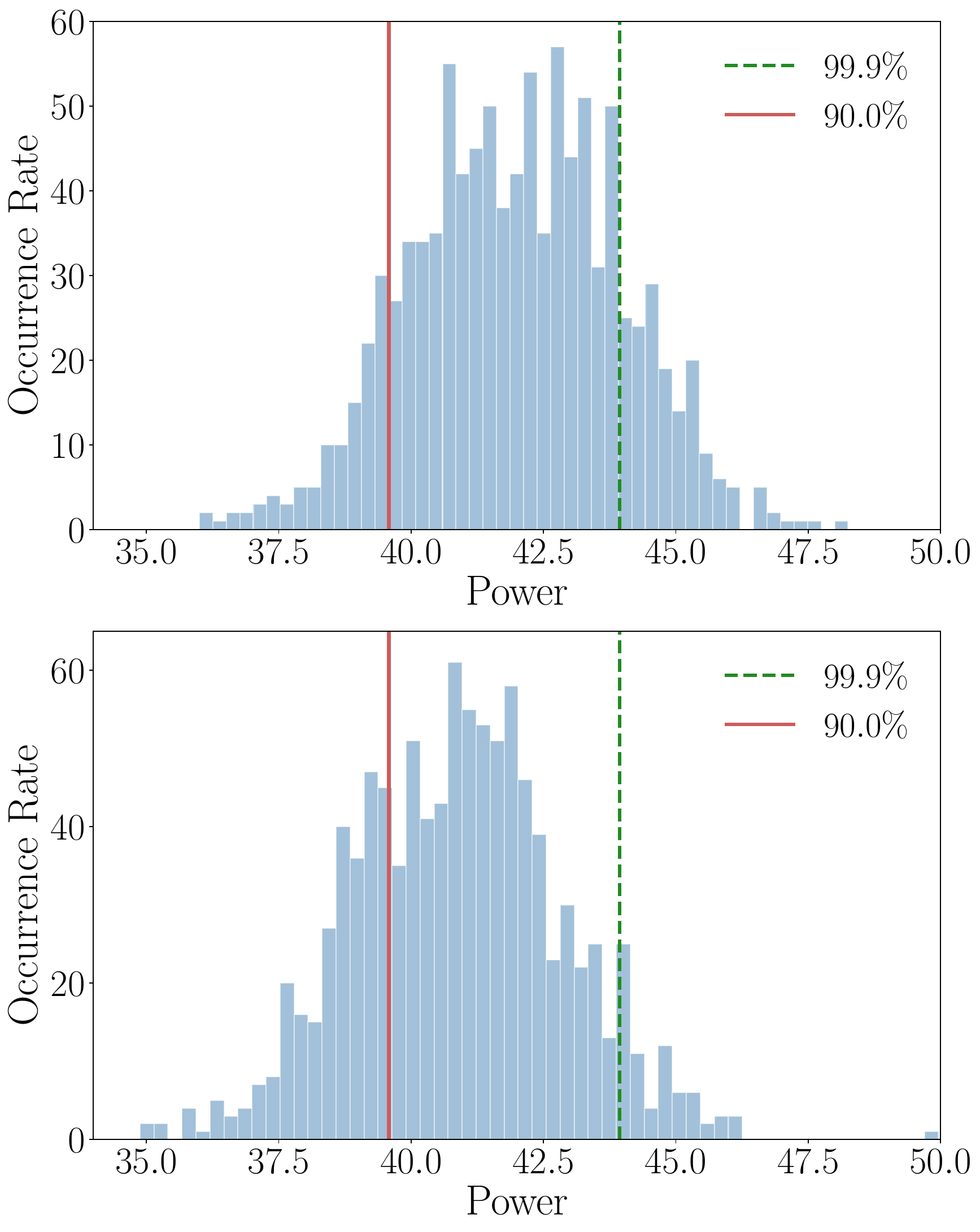}
    \vspace*{-0.5cm}
    \caption{Histograms showing the power distribution of the top (strongest) candidate from 1000 data reduction iterations (conducted separately from the main searches) and $Z^{2}_{3}$ searches of \textit{XMM-Newton} ObsID 0724060801, in the frequency range $1.20 - 1.30$ Hz (top) and $1.21 - 1.31$ Hz (bottom). Here, the occurrence rate is the total number of times a candidate appears within each power bin across the 1000 iterations.}
    \label{fig:DetSig}
\end{figure}

The lower panels of Figure \ref{fig:ZMid}, show the power distributions of the top candidate from each data reduction iteration from the $Z^{2}_{2}$ (left) and $Z^{2}_{3}$ (right) searches. Respectively, the solid and dashed lines show the $99.9 \%$ statistical detection threshold (confirmed pulsations with \textsc{hendrics}) and the $90 \%$ statistical detection threshold. In all cases, the top candidate is $1.269$ Hz. However, the number of statistically confirmed pulsations differs drastically between the two searches due to the randomisation within the data reduction procedure and the harmonic content used for the search. The $Z^{2}_{2}$ uses a lower harmonic content than the $Z^{2}_{3}$ search, so the power corresponding to the 99.9 per cent detection threshold, as determined by \textsc{hendrics}, is lower than that of the $Z^{2}_{3}$ search. Respectively, for the $Z^{2}_{2}$ and $Z^{2}_{3}$ searches, these values are $37.96$ and $43.97$. The number of candidates satisfying these statistical detection limits is $82.9~\%$ and $2.4~\%$ of the top candidates, respectively for the $Z^{2}_{2}$ and $Z^{2}_{3}$ searches. For the 90 per cent statistical detection threshold, these values become $34.2$, corresponding to $99.9~\%$ per cent of the top candidates for the $Z^{2}_{2}$ search and $39.61$, corresponding to $61.0~\%$ of the top candidates for the $Z^{2}_{3}$ search. These results show that had we only performed one data reduction iteration and $Z^{2}_{3}$ search (as was completed in our initial analysis in Section \ref{sx:Data}), we would have a $\approx 3$ per cent chance of finding a statistically confirmed pulsation at $1.27$ Hz. In contrast, if performing one data reduction iteration and $Z^{2}_{2}$ search, there is an $83$ per cent chance of returning a statistically confirmed pulsation. This further highlights the potential zealous presentation of marginal signals if the effects of the data reduction randomisation are ignored. Despite the power variations among the $Z^{2}_{2}$ (left) and $Z^{2}_{3}$ (right) searches, there is a remarkable consistency among the returned signals -- the returned frequency is always $1.269$ Hz (to three decimal places) and the folded pulse profiles are skewed sinusoids in each case. We demonstrate this with Figure \ref{fig:FreqSpec}, which shows the highest and lowest significance pulsation candidates from the $Z^{2}_{2}$ (rows one and two) and $Z^{2}_{3}$ (rows three and four) searches.

Finally, we present Figure \ref{fig:DetSig}, which shows the top candidates (again, always $1.269$ Hz) from one thousand data reduction iterations (conducted separately from the main searches) and searched using a $Z^{2}_{3}$ search in the $1.2 - 1.3$ Hz band (top) and the $1.21 - 1.31$ Hz (bottom) bands. Here, we see a surprising difference in the resulting power distributions, especially considering that the search band shifts by only $0.01$ Hz. In both cases, the 90 and 99.9 per cent statistical detection thresholds are $39.59$ and $43.94$, respectively. However, the percentage of candidates surpassing each threshold is vastly different. When searching the $1.2 - 1.3$ Hz band, we find $886 / 1000$ of the top candidates surpass the 90 per cent statistical detection threshold while $163 / 1000$ of the top candidates surpass the 99.9 per cent limit. These numbers become $888 / 1000$ and $62 / 1000$, respectively, when searching the $1.21 - 1.31$ Hz band (see Table~\ref{tb:searches} for details of the number of candidates meeting these statistical detection thresholds for all searches conducted within this Section). Therefore, it is evident that the strength of a pulsation signal depends on both the randomisation in the data reduction procedure and the search parameters. These dependencies, among others, are explored in detail in a forthcoming paper. Due to these variations, we determine that we are unable to assign a single power to the promising $1.269$ Hz candidate. Instead, we report search-dependent significances in Table \ref{tb:searches}, using the mean and standard deviation of the power distributions complied from each search. 

\begin{figure}
    \centering
    \includegraphics[width=\columnwidth]{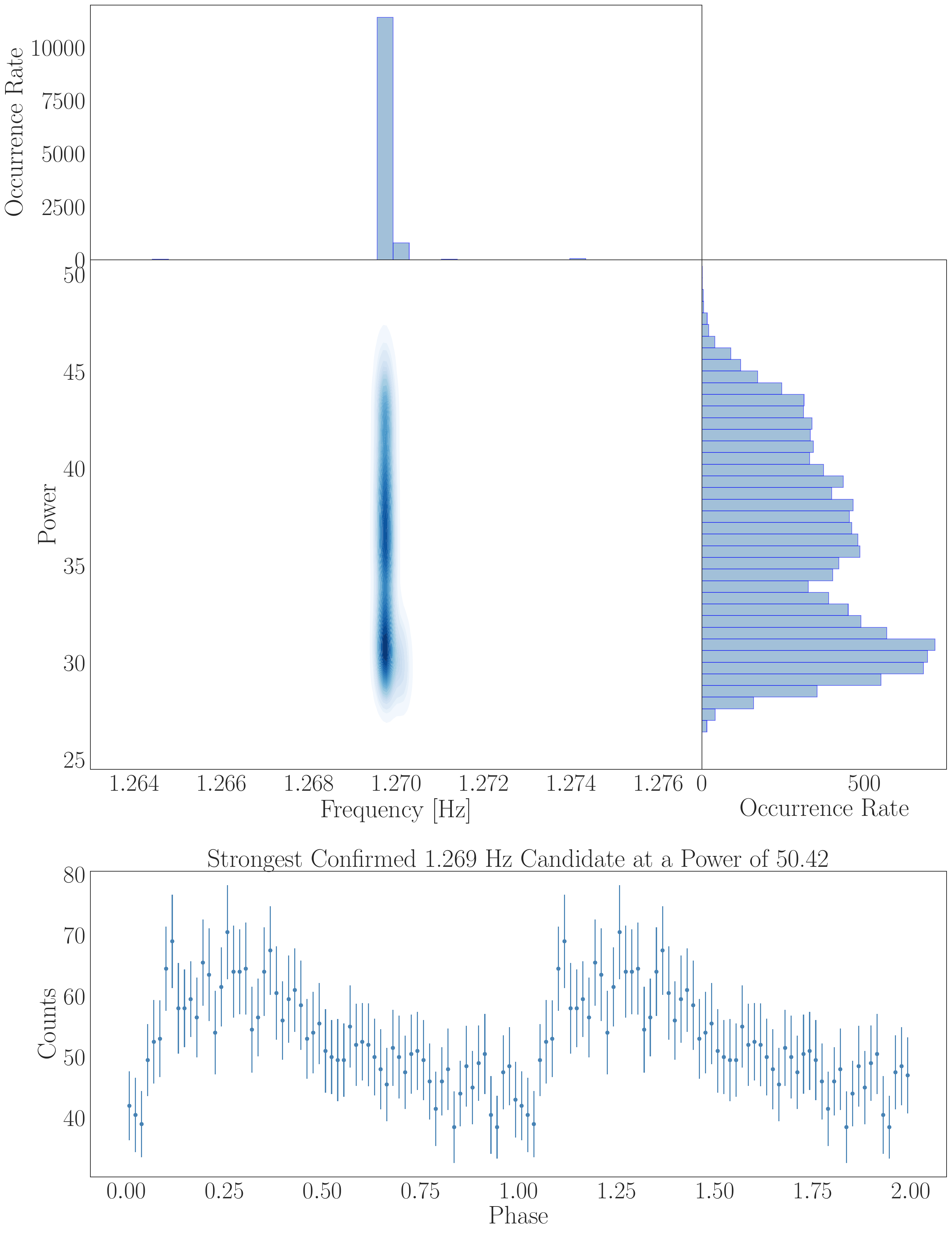}
    \vspace*{-0.35cm}
    \caption{Top: Histograms showing the candidate pulsation frequencies and corresponding powers returned by all candidates when conducting 2500 data reduction iterations and performing a $Z^{2}_{3}$ search of \textit{XMM-Newton} ObsID 0724060801, in the frequency range $1.26 - 1.28$ Hz. Bottom: The folded pulse profile of the strongest, statistically confirmed candidate found in data reduction iteration 150. Here, the occurrence rate is the total number of times a candidate appears within each frequency (top) or power (side) bin across the 2500 iterations.}
    \label{fig:ZNarrow}
\end{figure}

\subsubsection{Narrow Band Searches}
We show the results of the narrow band ($1.26 - 1.28$ Hz) $Z^{2}_{3}$ search results in Figure \ref{fig:ZNarrow}, where we see the frequency and power distribution of all returned candidates in the top panel and an example folded pulse profile from the statistically confirmed pulsation detection in data reduction iteration $150$ in the bottom panel. When narrowing the frequency band used to conduct the searches and requesting a higher harmonic content, the search results are clear -- there is one dominant pulsation candidate at $\sim 1.269$ Hz. Despite the prominence of this signal, we still cannot be certain of its power. As shown in the right-hand histogram in Figure \ref{fig:ZNarrow} and Table \ref{tb:searches}, the power of this signal ranges from $\sim 26 - 50$. Using the mean and standard deviation of the resulting power distribution, we determine that our narrow band $Z^{2}_{3}$ search finds a $1.269$ Hz pulsation candidate with a power of $ 36.2 \pm 4.8$ ($1~\sigma$). While these results support the presence of a $1.269$ Hz candidate in this data set, they also demonstrate a potential risk of narrowing the frequency band too far -- an artificially strong candidate. Simply put, when searching a narrow frequency range, there are fewer n-trials. Therefore, the search algorithm is unaware of any signals outside the specified range, so that a candidate that appears prominent compared to others within that small range is returned with a power that reflects its strength relative to the limited number of other trailed frequencies. Over a larger frequency range, that candidate may no longer be distinguishable from the surrounding frequencies. The situation is further complicated in the case of marginal detections because they are naturally more challenging to distinguish from the surrounding frequencies. However, based on the search results presented throughout this paper, we suggest that even marginal, transient candidates with powers $<35$ are distinguishable, and can be found by searching for a signal that does not disappear when conducting multiple searches with different parameters, rather than focusing simply on the strength of the signal. 

\subsection{Signal Verification}
\label{sx:verify}

\subsubsection{Additional Observations}
The best way to confirm that J1325 is a PULX is to have a corroborating pulsation detection from another observation. Thus far, we have performed multiple data reduction iterations and pulsation searches on \textit{XMM-Newton} ObsID 0724060801 in the $1 - 10$ keV band. However, as noted in Section \ref{sx:Data} and seen in Figure \ref{fig:Long_Term_Lightcurve} another \textit{XMM-Newton} observation, ObsID 0724060701, captured J1325 during its 2014 outburst, roughly one month earlier. As such, we apply the search procedure shown in Figure \ref{fig:Flow} to ObsID 0724060701, utilising the $1 - 10$ keV band to try and corroborate our findings. We summarise the results below and provide the corresponding figures as supplementary material.  

We perform an initial $n=100$ data reductions and stage one searches to see whether a candidate close to $1.27$ Hz emerges. Informed by the analysis presented throughout this section, we chose to perform 100 data reduction iterations, initially, as this is sufficient to obtain frequency and power distributions that are representative of those obtained by a higher number of iterations, while minimising the computational time. Should any viable pulsation candidates emerge, we would conduct further iterations. However, there are no strong candidates close to $1.27$ Hz in either the standard acceleration search or the acceleration search with interbinning and a distinct lack of signals between $1 - 3$ Hz in both cases. The only signal found by both stage one searches is at $\sim 4.9$ Hz, which has the highest occurrence rate in both cases. Both acceleration searches also return a cluster of signals at $\lesssim 1$ Hz. The interbinned search exhibits hot spots between $2.5 - 3$ Hz and a peak at $\sim 2$ Hz, but neither appears in the standard acceleration search. These results suggest that no corroborating signal is present in this earlier \textit{XMM-Newton} observation, which could be because the pulsations from J1325 are transient. Alternatively, it may be because this observation occurred earlier in the outburst, when the source was both softer and fainter, making pulsations even more challenging to detect. Figures summarising the stage one searches of ObsID 0724060701 are provided as online supporting material.

For completeness, we continue following the search procedure in Figure \ref{fig:Flow} and conduct preliminary stage two searches in the following bands: $4.75 - 5.25$, $2.5 - 3$ and $0.5 - 1.0$ Hz. We additionally perform preliminary searches in the $1.02 - 1.52$ Hz band to verify the $1.27$ Hz signal is not present. We initially see evidence of a $\sim 0.6$ Hz signal as it has the highest occurrence rate in all three $0.5 - 1.0$ Hz band searches. However, the peaks fall into different frequency bins and do not show the same agreement as seen for the $1.27$ Hz signal in Figure \ref{fig:Prelim_Band1}. We also note that the top candidate is not always the signal with the highest occurrence rate and that the folded pulse profiles are not always sinusoidal. The same arguments allow us to discard the results from the $1.02 - 1.52$ Hz searches. In particular, we note the lack of a corroborating $1.27$ Hz signal and disagreement between the top candidates which range from $\sim 1.1$ Hz in the $Z^{2}_{2}$ search to $\sim 1.5$ Hz in the $Z^{2}_{3}$ search. The results from the search in the $2.5 - 3$ Hz band display more consistency, with each search identifying a top candidate at $\sim 2.6$ Hz. However, the lack of sinusoidal pulse profiles allows us to discard these candidates. Lastly, the $Z^{2}_{1}$ search in the $4.75 - 5.25$ Hz band returns a near perfect sinusoidal folded pulse profile at a frequency of $5.185$ Hz. Nonetheless, as the $Z^{2}_{2}$ and $Z^{2}_{3}$ searches do not find a supporting signal, so we discard this candidate and conclude that there are no viable pulsation candidates in \textit{XMM-Newton} ObsID 0724060701. Further details and figures summarising these search results are provided as online supporting material.


\subsubsection{Full Energy Band}
PULXs are known to exhibit harder-than-average X-ray spectra compared to the population of ULXs, and the PULXs themselves are spectrally harder when the intensity of the pulsation component increases. Therefore, one way to uncover weaker pulsation signals is to search in a harder X-ray band. We have applied this principle to the searches already presented in this section, by eliminating the softest part of \textit{XMM-Newton's} bandpass (i.e. using energies $> 1$ keV). Nonetheless, we look to verify our findings by applying our search procedure to the full \textit{XMM-Newton} bandpass ($0.2 - 10$ keV) to determine whether the $1.27$ Hz signal emerges. To the end, we conduct 100 data reduction iterations of \textit{XMM-Newton} ObsID 0724060801, following the reduction procedure described in Section \ref{sx:Data} to obtain events lists in the $0.2 - 10$ keV energy band, and subject these event lists to the search procedure shown in Figure \ref{fig:Flow}, with one difference -- we bypass the preliminary searches as we know the frequency we are attempting to verify. As shown in Table \ref{tb:searches}, the $1.27$ Hz signal emerges during the stage two searches, albeit at lower significance than in the $1 - 10$ keV band searches. However, the $1.27$ Hz signal does not appear often or with high powers in either of the acceleration searches or the wide band stage two search. This indicates a dilution of the signal by soft photons from an invariant component, making it challenging to consistently detect. 

\subsubsection{Harmonics}
Our search procedure makes use of $Z^{2}_{1}$, $Z^{2}_{2}$ and $Z^{2}_{3}$ searches. Respectively, these assess signals with one (also referred to as the fundamental frequency), two and three harmonics. Therefore, the results of a $Z^{2}_{1}$ search should return all reasonable candidates for the fundamental frequency. Similarly, the results $Z^{2}_{2}$ and $Z^{2}_{3}$ searches should return signals for which the second and third harmonics are present. Therefore, all three versions of the $Z^{2}$ search should return the fundamental frequency. Indeed the preliminary searches (Figure \ref{fig:Prelim_Band1}) and our mid-band searches (Figure \ref{fig:ZMid}) all return our 1.27 Hz candidate. To first order, this supports a 1.27 Hz pulsation signal in the data, but there is a caveat -- the strength of harmonic signals decrease with each integer step. For pulsation candidates detected at marginal significance, this may mean that the harmonics of the signal are too weak to enable a confident detection of the fundamental frequency in searches demanding a higher harmonic content. This may explain the stark difference between the number of $1.27$~Hz candidates surpassing the $99.9~\%$ statistical detection limit in our mid-band $Z^{2}_{2}$ and $Z^{2}_{3}$ searches (see Figure \ref{fig:ZMid} and Section \ref{sx:mid}). 

While the results of our $Z^{2}_{2}$ and $Z^{2}_{3}$ searches imply that the second and third harmonics of our $1.269$ Hz signal are present within the data, we choose to verify this by searching for them manually. Assuming integer steps, the second and third harmonics are $2.538$ and $3.807$ Hz, respectively. To the end, we conduct 100 data reduction iterations of \textit{XMM-Newton} ObsID 0724060801, following the reduction procedure described in Section \ref{sx:Data} to obtain events lists in the $1.0 - 10$ keV energy band, and subject these event lists to targeted $Z^{2}_{1}$ searches in 0.25 Hz bands around the harmonic frequencies. Here, using the $1 - 10$ keV event lists should enhance the signal (see previous sub-section). As these frequencies are harmonics of the $1.27$ Hz signal, their second and third harmonics are equivalent to the fourth and sixth, and sixth and ninth harmonics of the fundamental frequency, respectively, and are likely too weak to detect. Therefore, searches with higher harmonic contents are not necessary here. The results of these searches are summarised in Table \ref{tb:searches}, which shows that the second and third harmonics are weak but present, increasing the likelihood that $1.27$ Hz is a genuine signal.

\section{Discussion}
\label{sx:Discuss}
Understanding what fraction of the ULX population hosts NS accretors has been a primary aim within the field since the discovery of the first PULX \citep{Bachetti2014}. This discovery challenged previous hypotheses, which suggested that accretion onto intermediate-mass black holes or super-Eddington accretion onto stellar-mass black holes powered all ULXs, and instead demanded models for super-Eddington accretion in the presence of a NS. Since then, a further seven extragalactic PULXs have been confirmed \citep{Furst2016, Israel2017, Carpano2018, Sathyaprakash2019, RodriguezCastillio2020, Pintore2025, Ducci2025}, in addition to some reasonable NS ULX candidates (see e.g. \citealt{Brightman2018, Gurpide2021, Quintin2021}). Here, we have uncovered evidence that the transient ULX J1325 hosts a NS accretor, as indicated by the robust (i.e. resilient to the type of pulsation search and parameters) detection of coherent X-ray pulsations at 1.27 Hz, albeit at marginal significance (i.e. typically, but not always, appearing below the statistical detection threshold within \textsc{hendrics}). We find the coherent X-ray pulsations in one of two \textit{XMM-Newton} observations taken during the source's only observed major outburst in 2014. However, we do not have the confirmation of a second, independent detection of this signal. Therefore, we cannot definitively claim a pulsating nature for J1325; this is a downside of observing transient sources with limited datasets.  Based on the presented pulsation detections, we consider J1325 a good but unconfirmed PULX candidate and discuss whether the other properties of the source support the preliminary PULX classification in the rest of this section. 

\subsection{Comparison of Properties to PULX Population}
J1325 is an understudied transient ULX, identifiable as a clear X-ray point source for $\sim 8$ months in 2014, during its only major recorded outburst (see Figure~\ref{fig:Long_Term_Lightcurve}). We detect coherent X-ray pulsations at a frequency of $1.27$~Hz in one \textit{XMM-Newton} observation during this outburst (ObsID 0724060801), at a marginal significance (see Section~\ref{sx:Pulse} and Table~\ref{tb:searches}). For our $1.27$~Hz candidate, we measure a pulsed fraction, PF $\approx 15 - 17~\%$, and $\dot{f} \sim 4 \times 10^{-9}$ Hz~s$^{-1}$. The X-ray spectrum of J1325 in the same observation in which we detect the candidate $1.27$~Hz pulsations is relatively soft and well described by an absorbed power law with photon index of $\Gamma \sim 2.7$. In addition to the 2014 outburst, J1325 briefly re-brightened in 2017 and 2024, but was not observed to cross the ULX luminosity threshold of $10^{39}$ erg~s$^{-1}$. We compare these properties of J1325 with the broader population of PULXs here. 

There are only a handful of detections of pulsations in ULXs (see e.g. Table 2 in \citealt{King2023} for a summary), the majority of which exhibit coherent X-ray pulsations on periods of the order of a few seconds. For example, M82 X-2 and NGC 5907 ULX-1, respectively, have measured pulsation periods of 1.4 and 1.1 seconds \citep{Bachetti2014, Bachetti2020, Israel2017, Fuerst2023}. However, the measured spin periods from persistent PULXs range from NGC 7793 P13 with a period of 0.4 seconds \citep{Furst2016} to NGC 300 ULX-1 with a period of 126 seconds \citep{Vasilopoulos2018}. Our marginal pulsation detections from J1325 fall within this range of measured spin periods at 0.787 seconds, making it consistent with the existing population of persistent PULXs. As shown in Table \ref{tb:searches}, the measured pulsed fraction from our \textsc{hendrics} searches changes depending on the type of pulsation search and the parameters used. Despite this, our measured pulsed fractions are remarkably consistent (and generally within tolerance of one another), falling within the range, PF~$\approx 15 - 17~\%$  in the $1 - 10$ keV band for our $1.27$~Hz candidate. This measurement is consistent with those from the population of persistent PULXs, although it is pertinent to note that these measurements were not all taken from the same instrument or energy band). Finally, for our $1.27$~Hz candidate we consistently measure $\dot{f} \sim 4 \times 10^{-9}$ Hz~s$^{-1}$ regardless of the search parameters (see Table \ref{tb:searches}), which is consistent with the range of reported $\dot{f}$ measurements from persistent PULXs (see e.g. \citealt{King2023}). While some of the persistent PULXs have negative $\dot{f}$ (e.g. NGC 5907 ULX-1 and M51 ULX-7), positive $\dot{f}$ measurements are not uncommon among PULXs and are expected from an accretion-driven torque. The positive $\dot{f}$ measurement could even indicate that J1325 is very close to spin-equilibrium, as is suggested for the persistent PULX M82 X-2 \citep{Bachetti2022}.

Perhaps the most intriguing finding from this study is the soft X-ray spectrum exhibited by our candidate while actively pulsating (assuming, of course, that our $1.27$~Hz detection is real and originates from J1325). PULXs typically exhibit a harder-than-average X-ray spectrum when compared to the broader population of ULXs. A dominant hard and variable spectral component in the $0.3 - 10$ keV range is likely the cause of the spectral hardness (see e.g. \citealt{Walton2018, Gurpide2021}), placing PULXs in the hard ultraluminous regime when the pulsations are on. Indeed, \citet{Gurpide2021} explored the spectral and timing properties of a sample of ULXs, including PULXs, finding that the PULXs are the spectrally hardest sources in their sample. The authors also showed that among their sample of known PULXs, those with higher pulsed fractions were spectrally harder (e.g. NGC 7793 P13: PF~$\sim 20~\%$; \citealt{Israel2017b} and M51 ULX-7: PF~$\sim 12~\%$ \citealt{RodriguezCastillio2020}), while those with lower pulsed fractions were slightly softer (e.g. NGC 1313 X-2: PF~$\sim 5~\%$ \citealt{Sathyaprakash2019}). This analysis suggested that spectral hardness could be an indicator for good PULX candidates. However, as seen in Figure \ref{fig:Spectra}, J1325 contests this, with its spectrum peaking at $\approx 1$ keV, which is more representative of the soft ultraluminous state. As discussed by \citet{Kobayashi2023} (and references therein), as the intensity of the pulsating component in NGC 300 X-1 decreases, a stable, thermal component peaking below $\sim 1$ keV becomes more apparent above the diminishing, variable harder continuum. Our stronger detection of pulsations in the 1-10 keV band than in the wider 0.2-10 keV band suggests similar behaviour -- the pulsed signal may be diluted by a stable soft component -- with the harder spectral component becoming more dominant at higher luminosities (see Table~\ref{tb:double}). However, we emphasise that this harder component is significantly softer than is typically seen in PULXs, at $\sim 1.1$ keV compared to $\gtrsim 2$ keV for the same underlying spectral model and similar pulse fractions \citep{Gurpide2021}.

The spectral softness of J1325 is similar to several of the soft ultraluminous sources studied by \cite{Gurpide2021}, notably Ho II X-1, NGC 5408 X-1 and NGC 55 ULX. The canonical interpretation of this softness in supercritical accretion models is that they are viewed at a relatively high inclination (see e.g. \citealt{Kaaret2017}), which is supported by the detection of dips in the light curve of NGC 55 ULX \citep{Stobbart2004}. However, at high inclinations, the central regions of the flow should be largely obscured by the photosphere, which in turn, would reprocess the emission from the central part of the flow responsible for the pulsations, effectively washing out any coherent signals. This effect makes it challenging to interpret the pulsation as real. However, there are possible explanations; for instance, we may be viewing the source such that we see the pulsation signal reflected from the far wall of the funnel structure. An alternative explanation is that inclination is not the dominant driver of ULX properties. Instead, we may be directly viewing the accretion curtain precessing to produce pulsations, as suggested by \citet{Mushtukov2017}. In this case, either the viewing angle is shallower (thus hiding the hardest emission) or the precessing region is larger and cooler than for other PULXs.  Whatever the solution, if the pulsation is real, the relative softness of J1325 requires an adjustment to previous models for PULXs. It also has implications for identifying PULX candidates -- simply looking for spectrally hard objects may overlook some fraction of PULXs that appear spectrally softer.

A further characteristic that J1325 has in common with other PULXs is that it displays high amplitude long-term variability, including periods of apparent quiescence (e.g. \citealt{Fuerst2023}). Initially, this type of long-term variability was associated with the propeller effect \citep{Tsygankov2016}, which led to searches for new PULX candidates on this basis \citep{Earnshaw2018,Song2020}. However, subsequent results showed that at least some of the long-term variability of PULXs is due to super-orbital periods on $\sim 50-100$ day timescales, likely due to precession of the accretion flow (e.g. \citealt{Brightman2019}). Indeed, there is now a suggestion that the propeller effect may predominantly result in a spectral softening of the source \citep{Middleton2023}. \citet{Khan2022} discussed whether precession in ULXs that are geometrically beamed could cause high amplitude X-ray variability. However, in this scenario we would expect regular detections of the PULXs on the timescale of their precession, as is seen in several such objects (e.g. 55-60 days in M82 X-2, \citealt{Kong2016,Brightman2019}; 78 days in NGC 5907 ULX, \citealt{Walton2016}). This is not the case for J1325, which instead has only been observed as a ULX in a single $\sim 8$ month outburst in 2014, with two subsequent less luminous detections in 2017 and 2024, over a baseline of $\sim 25$ years (see Figure~\ref{fig:Long_Term_Lightcurve}).

This apparent transient behaviour is not uncommon for ULXs (see e.g. \citealt{Middleton2012,Middleton2013} for studies of example transient ULXs in M31, and \citealt{Brightman2023} for discussion of a sample of five transient ULXs). However, within this transient subset of ULXs, there are a handful of known pulsating transients with outbursts that reach the ultraluminous regime (see Table 2 of \citealt{King2023}), some of which could be Be X-ray binary systems. This class includes the well-studied Galactic PULX, \textit{Swift} J0243.6$+$6124 \citep{Doroshenko2018,Tsygankov2018}. Similar to J1325, the outbursts of these PULX transients barely exceed the ULX threshold ($L_{\rm max} \sim 2-3 \times 10^{39} \rm~erg~s^{-1}$). Typically, these transient Be-PULXs spin down in between outbursts and rapidly spin up (positive $\dot{f}$) when entering a so-called superoutburst, when their luminosity exceeds $10^{39} \rm~erg~s^{-1}$. This scenario is consistent with our measurements for J1325. However, their pulse periods are generally longer at $\gtrsim 10$~s and, like the persistent PULXs, these sources tend to be spectrally hard. For example, \citet{Trudolyubov2007a} determined the transient PULX NGC 2403 ULX is spectrally hard, with a best-fitting photon power law index in the range $\Gamma \sim 0.9 - 1.2$. Similarly, \citet{Chandra2020} fit the broadband $0.5 - 50$ keV spectra of RX J0209.6-7427 with a cut-off power law model, stating that the spectrum during the outburst was relatively hard ($\Gamma \sim 2.2$ with a partial covering absorption fraction of $\sim 0.87$ and $N_{\rm{H}}=$ 16 $\times 10^{22}$), as did an investigation of the spectral and timing behaviour of the first Galactic PULX, \textit{Swift} J0243.6$+$6124 ($\Gamma \lesssim 1.25$; see \citealt{Jaisawal2018}). Overall, J1325 is generally consistent with the existing population of PULXs, but only partly aligns with each of the persistent and transient subsets. Therefore, it is clear that a better understanding of the mechanisms driving the emission contributing to the observed X-ray spectra in PULXs is required to understand the full range of PULX behaviours.

\subsection{Reproducibility of Timing Signals}
We have presented a thorough pulsation search procedure that considers the reproducibility of a signal as an indicator of a good pulsation candidate, rather than basing conclusions solely on the signal strength resulting from a single data reduction and search. Our approach is based on the premise that a real signal will appear in the search results regardless of how the pulsation search is conducted, and accounts for the intrinsic randomisation within \textsc{xmm-sas}. Our search procedure employs two stages to identify candidates that occur consistently and discards any candidates that appear only under certain conditions. As shown in Table \ref{tb:searches}, our favoured $1.27$~Hz candidate is found in all $1 - 10$ keV searches that encompass this frequency ~\footnote{The $1.27$~Hz candidate is found in the mid-band Z$^{2}_{1}$ search, but it does not fall into the 1~$\sigma$ distribution. See also Figure~\ref{fig:ZMid}.}. However, we have not been able to confirm its astrophysical origin by detecting a corroborating signal in a second observation (see Section \ref{sx:verify}).  Hence the possibility remains that this signal may be a false positive induced by a random alignment of photon arrival times (as must be found if sufficient numbers of datasets are searched), rather than a true pulsation signal from J1325.

Nevertheless, as discussed above, this signal displays all the hallmarks of a physical pulsation signal from a PULX, including a sinusoidal pulsed profile. However, determining the true significance of our candidate pulsation is complicated, both by the intrinsic randomisation employed during the generation of clean event lists and the way the search is conducted. This has led to our $1.27$~Hz signal appearing at powers between 15.3 and 50.4 over thousands of data reductions and searches. A key point to reiterate is that unless the random seed variable within \textsc{XMM-SAS} is defined, the event list and therefore science results will differ each time. Fixing the random seed variable to a known integer each time the data reduction is performed removes these variations. However, it is highly non-trivial to determine what that integer value should be when each value produces a slightly different cleaned event list. 

At the lower end of this power range, signals are readily discarded due to their low strengths as they are barely distinguished from the surrounding frequencies. At the top end, the signals are much more compelling and warrant further consideration. Our results show that a single candidate frequency can emerge from the same data, reduced and searched using exactly the same command line arguments every time, at significantly different strengths. It is not only the data reduction randomisation introducing variability into the results, either. As shown in Figure \ref{fig:DetSig}, the parameters of the search also change the resulting power of the signal. Therefore, candidates emerging from pulsation searches of \textit{XMM-Newton} data with tools such as \textsc{hendrics} must be assigned a search-dependent power built up over a large number of searches to get a realistic idea of the believability of the detection. As shown throughout Section \ref{sx:Pulse}, our repetitive search method works with these known variations and can robustly determine whether a signal, even a marginal one, is present in the data. This will aid the discovery of new PULXs without needing to limit the sample to the most luminous ULXs or the ones with the hardest X-ray spectra.  It will also reveal whether past published detections at moderate significance (e.g. \citealt{Sathyaprakash2019, Quintin2021, Pintore2025}) are false positives or believable detections (Pagan et al., in prep.).

We note that the lack of reproducibility due to varying source event list selections has the potential to have a fundamental impact on both existing and future results, within the ULX field and beyond. Simply put, reducing the same set of ODFs thousands of times is not standard practice. While one might work through some reduction steps a few times to tweak source regions or improve the removal of background flares, this is not sufficient to reveal the true extent of the variations in the result. For high-significance results, these variations will still be present, but should not be a concern -- a high-significance detection is extremely unlikely to prove a false positive due to this effect. However, the variations are important for marginal detections as they might appear significant in some iterations and not in others. While we have demonstrated this is important for pulsation searches, this effect may also impinge on other analyses where marginal detections are important (e.g. source detection), and we encourage other teams to investigate this. We also note that while our investigation and results are specific to \textit{XMM-Newton} data reduced with \textsc{XMM-SAS}, data from other instruments are not immune to these same issues if the reduction method uses random number generation.

The core implication of our work is that for ULXs, there may be PULX candidates in the literature with pulsation signals whose strength is inaccurately reported, or even a false positive detection; and other potential detections that may have been missed. These inaccuracies hinder our ability to constrain the relative populations of NS and BH ULXs, and will have a knock-on effect on our understanding of accretion and pulsation mechanisms in ULXs and, more generally, the characteristics attributed to a NS ULX. Overall, our findings emphasize that corroborating pulsation detections, either by detection in different exposures or by different instruments, is of the highest value. 

\section{Conclusion}
\label{sx:conc}
In this paper, we have detected a plausible pulsation from the transient ULX J1325, indicative that it hosts an accreting NS rotating at a frequency of $1.269$ Hz. We have compared its properties to those of the known PULXs, noting that the pulse characteristics and its sinusoidal shape are consistent with other PULXs. Its long-term X-ray variability is not consistent with the more persistent well-known PULXs, however there are some known transient PULXs with similar variability behaviour. While the pulsations from J1325 appear stronger at harder energies (above 1 keV), similar to other PULXs, its overall spectrum is however unusually soft for a PULX. In the absence of a second, confirmatory detection of the pulsation in a separate observation, this may either indicate that this is a false positive detection, or that PULXs may have a wider range of spectral hardness (and underlying physics) than previously thought.

In addition, we present a robust method for pulsation searches that uses the inherent randomisation within the {\it XMM-Newton} data reduction and the premise that a genuine candidate pulsation signal will appear in the results regardless of how we search the data. We consistently identify coherent X-ray pulsations at $1.269$ Hz in \textit{XMM-Newton} ObsID 0724060801. This candidate displays a sinusoidal folded pulse profile and possesses second and third harmonics. However, the significance of this candidate is uncertain and complicated by the intrinsic randomisation, emerging with powers in the range of $15.3 - 50.4$.  Nevertheless, the persistence of this detection, with consistent parameters, throughout our analyses lends it credibility.  The variability of the significance of the signal has implications for past analyses of ULX data, that were frequently based on a single data reduction, meaning that both some weak and/or false positives may have been reported as strong detections in the literature; and some plausible signals may have been missed.  We will address this in future work.

\section*{Acknowledgements}
TPR and AHK acknowledge support from the Science and Technology Facilities Council (STFC) as part of the consolidated grant award ST/X001075/1.
DJW acknowledges support from STFC as part of the individual grant award ST/Y001060/1. 
This research has made use of software and data provided by the High Energy Astrophysics Science Archive Research Center (HEASARC) and SAO/NASA Astrophysics Data System. This research has made use of data obtained from the \textit{Chandra} Data Archive provided by the \textit{Chandra} X-ray Center (CXC). The work also makes use of data provided by the \textit{XMM-Newton} Science Archive and the \textit{XMM-Newton} science analysis software made publicly available by the European Space Agency. This work made use of data supplied by the UK \textit{Swift} Science Data Centre at the University of Leicester. Some figures presented in this work were created using \textsc{seaborn}, a statistical data visualisation package for \textsc{python} \citep{seaborn}. This research has made use of the XRT Data Analysis Software (XRTDAS) developed under the responsibility of the ASI Science Data Center (ASDC), Italy.

\section*{Data Availability}
The data used in this research are publicly available from either the \textit{Chandra} Data Archive provided by the \textit{Chandra} X-ray Center (CXC), the \textit{XMM-Newton} Science Archive provided by the European Space Agency or the High Energy Astrophysics Science Archive Research Center (HEASARC) website. 


\bibliographystyle{mnras}
\bibliography{ULX_Refs} 

\begin{footnotesize}
\onecolumn
\setlength\LTcapwidth{\textwidth}
\begin{longtable}[c]{p{2.6cm} | p{1.1cm} p{2.05cm} p{0.8cm} p{0.8cm} p{1.5cm} p{0.7cm} p{0.7cm} p{1.9cm} lp{1.0cm}}
\caption{\label{tb:searches} A summary of all pulsation searches of \textit{XMM-Newton} ObsID 0724060801 conducted within Section \ref{sx:Pulse}. The search results are present in the order in which they are conducted (see also Figure \ref{fig:Flow}). For each search type, we present results for a subset of the returned frequencies comprising $\geq 68~\%$ ($ 1~\sigma$) of the candidate distribution. The exception to this is the acceleration search with interbinning for which we present the ten highest occurrence rate candidates. The results from each search are presented in order of mean power (strength), highest to lowest. The number of data reduction iterations is $n$ and the total number of search results is $R$. Note that dashes indicate that the measurement was not possible while zero entries are genuine measurements.}\\
\toprule
 \ Search Details & Frequency & Occurrence Rate & No. $\geq$ & No. $\geq$ & Mean Power & Highest & Lowest & Mean Frequency & Mean Pulsed \\
\vspace*{+0.1cm}
and Parameters &[Hz] & (All | Top) & $90~\%$ & $99.9~\%$ & $\pm 1~\sigma$ & Power & Power & Derivative & Fraction [$\%$] \\
\hline \\
\vspace*{+0.1cm}
 & 0.281 & 9.83~$\%$ | 36.9$~\%$  & -- & -- & 32.38 $\pm$ 1.62 & 38.47 & 28.52 & 1.67 $\times 10 ^{-8}$ & --\\
\vspace*{+0.1cm}
Standard Acceleration & 1.269 & 18.7~$\%$ | 52.2$~\%$ & -- & -- &  31.99 $\pm$ 1.80 & 39.63 & 28.51 & 4.18 $\times 10 ^{-9}$ & -- \\
\vspace*{+0.1cm}
$0.01 - 5$ Hz \ $ 1 - 10$ keV & 0.051 & 8.40~$\%$ | 4.20$~\%$ & -- & -- & 30.53 $\pm$ 1.26 & 35.68 & 28.51 & -1.38 $\times 10 ^{-7}$ & -- \\
\vspace*{+0.1cm} 
$n = 2500$ \ $R = 25382$ & 1.465 & 7.02~$\%$ | 1.44$~\%$  & -- & -- & 30.04 $\pm$ 1.09 & 35.84 & 28.52 & -7.11 $\times 10 ^{-8}$ & --\\
\vspace*{+0.1cm} 
& 2.823 & 9.85~$\%$ | 2.36$~\%$  & -- & -- & 30.02 $\pm$ 1.16 & 35.87 & 28.51 & -5.77 $\times 10 ^{-8}$ & --\\
\vspace*{+0.3cm}
& 1.976 & 14.0~$\%$ | 2.24$~\%$  & -- & -- & 29.92 $\pm$ 1.10 & 35.71 & 28.51 & 7.85 $\times 10 ^{-8}$ & --\\

\hline \\
\vspace*{+0.1cm}
& 0.876 & 0.62~$\%$ | 5.08$~\%$  & -- & -- & 38.64 $\pm$ 4.25 & 51.91 & 31.58 & 1.11 $\times 10 ^{-8}$ & --\\ 
\vspace*{+0.1cm}
& 4.171 & 1.22~$\%$ | 70.9$~\%$ & -- & -- & 37.02 $\pm$ 6.23 & 58.75 & 31.59 & 5.03 $\times 10 ^{-9}$ & -- \\
\vspace*{+0.1cm}
& 2.823 & 1.34~$\%$ | 0.36$~\%$  & -- & -- & 36.69$\pm$ 3.52 & 48.25 & 31.57 & -5.77 $\times 10 ^{-8}$ & -- \\
\vspace*{+0.1cm}
Acceleration with & 1.976 & 1.97~$\%$ | 0.08$~\%$  & -- & -- & 36.66 $\pm$ 2.91 & 47.36 & 31.59 & 7.91 $\times 10 ^{-8}$ & -- \\
\vspace*{+0.1cm}
Interbinning & 3.485 & 3.87~$\%$ | 5.76$~\%$  & -- & -- & 36.58 $\pm$ 3.49 & 50.50 & 31.58 & 1.48 $\times 10^{-7}$ & --  \\
\vspace*{+0.1cm}
$0.01 - 5$ Hz & 4.237 & 1.09~$\%$ | 7.80$~\%$  & -- & -- &36.26 $\pm$ 4.27 & 54.96 & 31.59 & 7.91 $\times 10 ^{-9}$ & --  \\
\vspace*{+0.1cm} 
$ 1 - 10$ keV & 2.021 & 1.17~$\%$ | 2.24$~\%$  & -- & -- & 35.80 $\pm$ 3.94 & 51.13 & 31.59 & 2.21 $\times 10 ^{-8}$ & -- \\
\vspace*{+0.1cm}
$n = 2500$ \hfill $R = 500026$ & 2.203 & 4.18~$\%$ | 0.00$~\%$ & -- & -- & 34.61 $\pm$ 2.37 & 46.04 & 31.59 & 1.20 $\times 10 ^{-7}$ & -- \\
\vspace*{+0.1cm}
& 0.281 & 2.31~$\%$ | 0.16$~\%$  & -- & -- & 35.41 $\pm$ 2.91 & 47.22 & 31.59 & 1.43 $\times 10 ^{-8}$ & -- \\
\vspace*{+0.3cm}
& 1.269 & 1.68~$\%$ | 0.08$~\%$ & -- & -- & 33.89 $\pm$ 1.91 & 44.50 & 31.59 & 3.92 $\times 10 ^{-9}$ & -- \\

\hline \\
\vspace*{+0.1cm}
Prelim Z$^{2}_{1}$& 1.269 & 20.0~$\%$ | 100 $~\%$  & 241 & 224 & 36.01 $\pm$ 1.76 & 42.37 & 30.85 & 4.82 $\times 10 ^{-9}$ & $14.69 \pm 1.34$ \\ 
\vspace*{+0.1cm}
 $1.02 - 1.52$ Hz& 1.411 & 14.6~$\%$ | 0.00 $~\%$  & -- & 0 & 24.11 $\pm$ 1.49 & 28.29 & 19.76 & -3.36 $\times 10 ^{-8}$ & $\leq 15.32$ \\ 
\vspace*{+0.1cm}
$1 - 10$ keV  & 1.476 & 18.0~$\%$ | 0.00 $~\%$  & -- & 0 & 21.03 $\pm$ 1.01 & 24.40 & 18.69 & -1.48 $\times 10 ^{-9}$ & $\leq 14.54$ \\ 
\vspace*{+0.3cm}
$n = 250$ \hfill $R = 1250$ & 1.082 & 17.6~$\%$ | 0.00 $~\%$  & -- & 0 & 20.92 $\pm$ 1.12 & 23.92 & 18.35 & 3.19 $\times 10 ^{-9}$ & $\leq 14.51$ \\

\hline \\
\vspace*{+0.1cm}
Prelim Z$^{2}_{2}$ & 1.269 & 41.4~$\%$ | 88.8 $~\%$  & -- & 0 & 31.45 $\pm$ 1.99 & 37.82 & 26.96 & 4.44 $\times 10 ^{-9}$ & $\leq 17.15$ \\ 
\vspace*{+0.1cm}
$1.02 - 1.52$ Hz & 1.333 & 29.9~$\%$ | 5.60 $~\%$  & -- & 0 & 29.71 $\pm$ 1.26 & 33.87 & 26.68 & 2.68 $\times 10 ^{-8}$ & $\leq 17.12$ \\
\vspace*{+0.1cm}
$1 - 10$ keV \\
\vspace*{+0.3cm}
$n = 250$ \hfill $R = 1250$ \\

\hline \\
\vspace*{+0.1cm}
Prelim Z$^{2}_{3}$ & 1.269 & 56.8~$\%$ | 100 $~\%$  & 101 & 5 & 39.30 $\pm$ 3.30 & 49.49 & 33.28 & 4.11 $\times 10 ^{-9}$ & $16.97 \pm 1.97$ \\ 
\vspace*{+0.1cm}
$1.02 - 1.52$ Hz & 1.333 & 12.3~$\%$ | 0.00 $~\%$  & -- & 0 & 35.77 $\pm$ 1.54 & 40.64 & 33.11 & 2.74 $\times 10 ^{-8}$ & $\leq 18.61$ \\
\vspace*{+0.1cm}
$1 - 10$ keV \\
\vspace*{+0.3cm}
$n = 250$ \hfill $R = 1250$ \\

\hline \\
\vspace*{+0.1cm}
& 1.838 & 14.6~$\%$ | 70.4$~\%$  & -- & 0 & 25.79 $\pm$ 1.67 & 29.92 & 18.17 & -3.33 $\times 10 ^{-8}$ & $\leq 14.61$ \\ 
\vspace*{+0.1cm}
& 2.121 & 12.1~$\%$ | 3.20$~\%$  & -- & 0 & 21.81 $\pm$ 1.11 & 25.82 & 19.44 & 2.41 $\times 10 ^{-8}$ & $\leq 14.58$ \\
\vspace*{+0.1cm}
Prelim Z$^{2}_{1}$ & 1.879 & 23.3~$\%$ | 1.20$~\%$  & -- & 0 & 21.36 $\pm$ 1.13 & 24.51 & 18.76 & 3.37 $\times 10 ^{-8}$ & $\leq 14.62$ \\
\vspace*{+0.1cm}
$1.75 - 2.25$ Hz & 2.231 & 8.56~$\%$ | 1.60$~\%$  & -- & 0 & 20.84 $\pm$ 1.14 & 24.21 & 17.77 & -1.44 $\times 10 ^{-8}$ & $\leq 14.75$ \\
\vspace*{+0.1cm}
$1 - 10$ keV  & 2.126 & 5.04~$\%$ | 0.00$~\%$  & -- & 0 & 20.83 $\pm$ 0.99 & 24.21 & 19.28 & 4.82 $\times 10 ^{-9}$ & $\leq 14.48$ \\
\vspace*{+0.1cm}
$n = 250$ \hfill $R = 1250$ & 1.904 & 6.56~$\%$ | 1.20$~\%$  & -- & 0 & 19.97 $\pm$ 0.98 & 22.61 & 18.35 & -4.82 $\times 10 ^{-9}$ & $\leq 14.56$ \\
\vspace*{+0.3cm}
& 2.006 & 4.96~$\%$ | 3.20$~\%$  & -- & 0 & 19.25 $\pm$ 0.71 & 20.76 & 17.84 & 7.44 $\times 10 ^{-9}$ & $\leq 14.56$ \\

\hline \\
\vspace*{+0.1cm}
& 1.837 & 20.0~$\%$ | 98.0$~\%$  & -- & 0 & 32.2 $\pm$ 1.75 & 37.39 & 27.54 & -3.37 $\times 10 ^{-8}$ & $\leq 16.73$ \\
\vspace*{+0.1cm}
& 2.086 & 8.32~$\%$ | 0.80$~\%$  & -- & 0 & 28.01 $\pm$ 1.06 & 31.11 & 26.21 & 2.38 $\times 10 ^{-9}$ & $\leq 16.73$ \\ 
\vspace*{+0.1cm}
Prelim Z$^{2}_{2}$ & 2.054 & 3.68~$\%$ | 0.00$~\%$  & -- & 0 & 27.67 $\pm$ 1.05 & 30.95 & 26.21 & 1.55 $\times 10 ^{-8}$ & $\leq 16.73$ \\

\caption[]{Continued}\\
\vspace*{+0.1cm}
$1.75 - 2.25$ Hz & 1.794 & 8.40~$\%$ | 0.00$~\%$  & -- & 0 & 27.65 $\pm$ 0.96 & 30.52 & 25.98 & 2.30 $\times 10 ^{-8}$ & $\leq 16.72$ \\ 
\vspace*{+0.1cm}
$1 - 10$ keV  & 1.823 & 14.1~$\%$ | 0.40$~\%$  & -- & 0 & 27.62 $\pm$ 0.97 & 31.29 & 25.67 & 8.02 $\times 10 ^{-9}$ & $\leq 16.71$ \\ 
\vspace*{+0.1cm}
$n = 250$ \hfill $R = 1250$ & 1.891 & 4.00~$\%$ | 0.40$~\%$  & -- & 0 & 27.41 $\pm$ 1.10 & 30.31 & 25.74 & 1.57 $\times 10 ^{-8}$ & $\leq 16.78$ \\ 
\vspace*{+0.1cm}
& 1.793 & 4.40~$\%$ | 0.00$~\%$  & -- & 0 & 27.36 $\pm$ 0.86 & 29.44 & 25.84 & 2.02 $\times 10 ^{-8}$ & $\leq 16.67$ \\ 
\vspace*{+0.3cm}
& 2.043 & 8.00~$\%$ | 0.00$~\%$  & -- & 0 & 27.34 $\pm$ 0.85 & 29.68 & 25.60 & -4.70 $\times 10 ^{-9}$ & $\leq 16.77$ \\

\hline \\
\vspace*{+0.1cm}
& 2.039 & 15.8~$\%$ | 71.6$~\%$  & 127 & 14 & 43.41 $\pm$ 3.09 & 48.88 & 33.97 & -3.35 $\times 10 ^{-8}$ & $16.90 \pm 1.96$ \\
\vspace*{+0.1cm}
Prelim Z$^{2}_{2}$ & 2.023 & 15.4~$\%$ | 10.8$~\%$  & -- & 0 & 36.99 $\pm$ 1.61 & 42.68 & 32.79 & -1.32 $\times 10 ^{-8}$ & $\leq 18.85$ \\
\vspace*{+0.1cm}
$1.75 - 2.25$ Hz & 1.838 & 10.6~$\%$ | 0.40$~\%$  & -- & 0 & 36.81 $\pm$ 1.68 & 41.65 & 33.01 & -3.37 $\times 10 ^{-8}$ & $\leq 18.81$ \\
\vspace*{+0.1cm}
$1 - 10$ keV & 1.823 & 19.2~$\%$ | 5.60$~\%$  & -- & 0 & 36.58 $\pm$ 1.47 & 41.60 & 32.77 & 7.33 $\times 10 ^{-9}$ & $\leq 18.77$ \\
\vspace*{+0.3cm}
$n = 250$ \hfill $R = 1250$ & 1.794 & 9.76~$\%$ | 2.00$~\%$  & -- & 0 & 36.24 $\pm$ 1.0 & 39.82 & 33.13 & 1.90 $\times 10 ^{-8}$ & $\leq 18.70$ \\

\hline \\
\vspace*{+0.1cm}
& 2.231 & 13.6~$\%$ | 36.4$~\%$  & -- & 0 & 20.16 $\pm$ 1.16 & 23.79 & 17.83 & -1.43 $\times 10 ^{-8}$ & $\leq 14.09$ \\ 
\vspace*{+0.1cm}
Prelim Z$^{2}_{1}$ & 2.354 & 26.8~$\%$ | 20.4$~\%$  & -- & 0 &  19.46 $\pm$ 0.95 & 24.35 & 17.54 & 1.44 $\times 10 ^{-8}$ & $\leq 14.09$ \\
\vspace*{+0.1cm}
$2.15 - 2.55$ Hz& 2.548 & 12.4~$\%$ | 11.6$~\%$  & -- & 0 &  19.42 $\pm$ 0.91 & 22.05 & 17.58 & -2.28 $\times 10 ^{-8}$ & $\leq 14.16$ \\
\vspace*{+0.1cm}
$1 -10$ keV & 2.440 & 7.84~$\%$ | 5.20$~\%$  & -- & 0 &  19.30 $\pm$ 0.93 & 22.25 & 17.75 & 2.19 $\times 10 ^{-8}$ & $\leq 14.05$ \\
\vspace*{+0.3cm}
$n = 250$ \hfill $R = 1250$ & 2.405 & 8.80~$\%$ | 5.20$~\%$  & -- & 0 &  19.06 $\pm$ 0.88 & 22.32 & 17.08 & -1.44 $\times 10 ^{-8}$ & $\leq 14.09$ \\

\hline \\
\vspace*{+0.1cm}
& 2.365 & 20.4~$\%$ | 62.8$~\%$  & -- & 0 & 29.60 $\pm$ 1.66 & 34.43 & 25.49 & 2.89 $\times 10 ^{-8}$ & $\leq 16.62$ \\ 
\vspace*{+0.1cm}
Prelim Z$^{2}_{2}$ & 2.405 & 11.6~$\%$ | 19.20$~\%$  & -- & 0 & 29.43$\pm$ 2.72 & 34.87 & 24.99 & -1.82 $\times 10 ^{-8}$ & $\leq 16.66$ \\ 
\vspace*{+0.1cm}
$2.15 - 2.55$ Hz & 2.175 & 17.4~$\%$ | 10.0$~\%$  & -- & 0 & 28.03 $\pm$ 1.39 & 32.33 & 25.25 & 2.01 $\times 10 ^{-8}$ & $\leq 16.63$ \\
\vspace*{+0.1cm}
$1 - 10$ keV & 2.547 & 5.36~$\%$ | 0.40$~\%$  & -- & 0 & 27.31 $\pm$ 1.10 & 30.26 & 24.87 & -2.02$\times 10 ^{-8}$ & $\leq 16.58$ \\ 
\vspace*{+0.3cm}
 $n = 250$ \hfill $R = 1250$ & 2.160 & 13.1~$\%$ | 3.20$~\%$  & -- & 0 & 27.02 $\pm$ 1.08 & 30.75 & 24.85 & -1.94$\times 10 ^{-8}$ & $\leq 16.63$ \\

\hline \\
\vspace*{+0.1cm}
& 2.485 & 12.6~$\%$ | 24.4$~\%$  & -- & 0 & 34.06 $\pm$ 1.58 & 39.08 & 30.43 & -3.08 $\times 10 ^{-8}$ & $\leq 18.12$ \\
\vspace*{+0.1cm}
& 2.215 & 4.64~$\%$ | 8.40$~\%$  & -- & 0 & 33.89 $\pm$ 1.43 & 38.88 & 31.66 & 2.20 $\times 10 ^{-8}$ & $\leq 18.08$ \\ 
Prelim Z$^{2}_{3}$ \vspace*{+0.1cm}
& 2.239 & 13.8~$\%$ | 20.4$~\%$  & -- & 0 & 33.53 $\pm$ 1.24 & 37.07 & 31.21 & 2.48 $\times 10 ^{-8}$ & $\leq 18.09$ \\ 
\vspace*{+0.1cm}
$2.15 - 2.55$ Hz & 2.405 & 17.7~$\%$ | 8.00$~\%$  & -- & 0 & 32.98 $\pm$ 1.10 & 36.99 & 31.02 & -1.18 $\times 10 ^{-8}$ & $\leq 18.10$ \\ 
\vspace*{+0.1cm}
$1 - 10$ keV & 2.365 & 8.08~$\%$ | 2.80$~\%$  & -- & 0 & 32.83 $\pm$ 0.85 & 35.75 & 30.86 & 2.57 $\times 10 ^{-8}$ & $\leq 18.01$ \\ 
\vspace*{+0.1cm}
$n = 250$ \hfill $R = 1250$ & 2.279 & 5.52~$\%$ | 4.00$~\%$  & -- & 0 & 32.80 $\pm$ 1.03 & 35.75 & 31.14 & 4.57 $\times 10 ^{-9}$ & $\leq 18.13$ \\ 
\vspace*{+0.3cm}
& 2.373 & 9.28~$\%$ | 5.20$~\%$  & -- & 0 & 32.67 $\pm$ 0.98 & 35.68 & 31.02 & -9.08 $\times 10 ^{-9}$ & $\leq 18.08$ \\

\hline \\
\vspace*{+0.1cm}
& 3.583 & 14.3~$\%$ | 34.8$~\%$  & -- & 0 &  22.66 $\pm$ 1.37 & 25.85 & 18.50 & -4.82 $\times 10 ^{-9}$ & $\leq 14.54$ \\
\vspace*{+0.1cm}
& 3.339 & 15.4~$\%$ | 27.2$~\%$  & -- & 0 & 21.48 $\pm$ 1.66 & 26.72 & 18.29 & -1.76 $\times 10 ^{-8}$ & $\leq 14.53$ \\ 
\vspace*{+0.1cm}
Prelim Z$^{2}_{1}$ & 3.699 & 12.2~$\%$ | 8.00$~\%$  & --  &  0 & 21.24 $\pm$ 1.21 & 24.75 & 18.10 & 2.41 $\times 10 ^{-8}$ & $\leq 14.56 $\\ 
\vspace*{+0.1cm}
$3.25 - 3.75$ Hz  & 3.696 & 9.44~$\%$ | 10.0$~\%$  & -- & 0 & 20.61 $\pm$ 1.22 & 25.02 & 18.58 & -3.37 $\times 10 ^{-8}$ & $\leq 14.53$\\ 
\vspace*{+0.1cm}
$ 1 - 10$ keV & 3.385 & 5.76~$\%$ | 0.00$~\%$  & -- & 0 & 20.26 $\pm$ 0.86 & 22.16 & 18.45 & -2.41 $\times 10 ^{-8}$ & $\leq 14.48$ \\ 
\vspace*{+0.1cm}
$n = 250$ \hfill $R = 1250$ & 3.292 & 8.08~$\%$ | 1.60$~\%$  & -- & 0 & 20.42 $\pm$ 0.98 & 22.98 & 18.27 & -2.30 $\times 10 ^{-8}$ & $\leq 14.56$ \\ 
\vspace*{+0.3cm}
& 3.394 & 6.56~$\%$ | 0.00$~\%$  & -- & 0 & 20.04 $\pm$ 0.83 & 22.17 & 18.50 & 9.04 $\times 10 ^{-9}$ & $\leq 14.49$ \\ 

\hline \\
\vspace*{+0.1cm}
& 3.339 & 31.4~$\%$ | 76.0$~\%$  & -- & 0 &  29.50 $\pm$ 1.85 & 35.19 & 25.46 & -1.85 $\times 10 ^{-8}$ & $\leq 16.70$ \\
\vspace*{+0.1cm}
Prelim Z$^{2}_{2}$ & 3.622 & 3.92~$\%$ | 6.00$~\%$  & -- & 0 &  28.76 $\pm$ 1.17 & 31.62 & 26.51 & -2.25 $\times 10 ^{-9}$ & $\leq 16.65$ \\
\vspace*{+0.1cm}
$3.25 - 3.75$ Hz & 3.583 & 16.7~$\%$ | 10.4$~\%$  & -- & 0 &  28.17 $\pm$ 1.31 & 32.78 & 25.52 & -2.46 $\times 10 ^{-8}$ & $\leq 16.74$ \\
\vspace*{+0.1cm}
$1 - 10$ keV  & 3.645 & 8.00~$\%$ | 1.60$~\%$  & -- & 0 &  27.69 $\pm$ 1.04 & 30.72 & 25.99 & 1.12 $\times 10 ^{-8}$ & $\leq 16.70$ \\
\vspace*{+0.1cm}
$n = 250$ \hfill $R = 1250$ & 3.484 & 5.68~$\%$ | 0.40$~\%$  & -- & 0 &  27.55 $\pm$ 0.99 & 30.24 & 25.85 & -1.65 $\times 10 ^{-8}$ & $\leq 16.63$ \\
\vspace*{+0.35cm}
& 3.699 & 4.24~$\%$ | 0.00$~\%$  & -- & 0 &  27.47 $\pm$ 0.97 & 30.18 & 26.20 & 2.47 $\times 10 ^{-8}$ & $\leq 16.71$ \\

\hline \\
\vspace*{+0.1cm}
& 3.359 & 15.4~$\%$ | 51.6$~\%$  & -- & 0 &  37.87 $\pm$ 1.93 & 42.21 & 32.92 & -1.91 $\times 10 ^{-8}$ & $\leq 18.54$ \\
\vspace*{+0.1cm}
Prelim Z$^{2}_{3}$ & 3.345 & 8.24~$\%$ | 8.40$~\%$  & -- & 0 &  36.25 $\pm$ 1.66 & 40.85 & 33.03 & -1.91 $\times 10 ^{-8}$ & $\leq 18.54$ \\

\caption[]{Continued}\\
\vspace*{+0.1cm}
 $3.25 - 3.75$ Hz & 3.273 & 17.6~$\%$ | 18.4~$\%$  & -- & 0 &  35.50 $\pm$ 1.51 & 39.86 & 32.36 & 2.85 $\times 10 ^{-8}$ & $\leq 18.65$ \\
\vspace*{+0.1cm}
$1 - 10$ keV & 3.580 & 11.0~$\%$ | 5.60$~\%$  & -- & 0 &  35.47 $\pm$ 1.38 & 38.58 & 32.81 & -2.98 $\times 10 ^{-8}$ & $\leq 18.59$ \\
$n = 250$ \hfill $R = 1250$ & 3.635 & 10.1~$\%$ | 2.00$~\%$  & -- & 0 &  35.26 $\pm$ 1.22 & 39.44 & 32.45 & 1.23 $\times 10 ^{-9}$ & $\leq 18.54$ \\
\vspace*{+0.1cm}
& 3.509 & 11.6~$\%$ | 6.40$~\%$  & -- & 0 &  35.07 $\pm$ 1.37 & 39.48 & 32.47 & -1.17 $\times 10 ^{-8}$ & $\leq 18.58$ \\

\hline \\
\vspace*{+0.1cm}
Wide Z$^{2}_{3}$ \hfill $1 - 10$ keV & 0.873 & 30.4 ~$\%$ | 58.2$~\%$  & 823 & 34 & 41.98 $\pm$ 3.14 & 51.87 & 34.28 & -2.56 $\times 10 ^{-8}$ & 17.23 $\pm$ 1.96 \\
\vspace*{+0.1cm} 
$0.77 - 1.77$ Hz & 1.269 & 45.0 ~$\%$ | 41.8$~\%$  & 410 & 4 & 40.29 $\pm$ 2.82 & 50.42 & 34.49 & 3.64 $\times 10 ^{-9}$ & 17.14 $\pm$ 1.96 \\
\vspace*{+0.3 cm} 
$n = 2500$ \hfill $R = 12500$ \\

\hline \\
\vspace*{+0.1cm}
Mid Z$^{2}_{1}$  \hfill  $1 - 10$ keV & 1.253 & 25.9~$\%$ | 87.1$~\%$  & -- & 0 & 20.73 $\pm$ 2.29 & 27.01 & 15.35 & -3.54 $\times 10 ^{-9}$ & $\leq$ 14.45 \\
\vspace*{+0.1cm}
$1.21 - 1.31$ Hz & 1.286 & 18.6~$\%$ | 0.03$~\%$  & -- & 0 & 18.77 $\pm$ 1.24 & 23.54 & 15.72 & -2.41 $\times 10 ^{-8}$ & $\leq$ 13.93 \\
\vspace*{+0.1cm}
$n = 2500$ \hfill $R = 12500$ & 1.214 & 18.4~$\%$ | 9.56$~\%$  & -- & 0 & 19.38 $\pm$ 1.50 & 24.10 & 15.49 & -1.44 $\times 10 ^{-8}$ & $\leq$ 14.10 \\
\vspace*{+0.3cm}
& 1.302 & 13.4~$\%$ | 0.00$~\%$  & -- & 0 & 17.35 $\pm$ 0.82 & 20.23 & 15.16 & -1.44 $\times 10 ^{-8}$ & $\leq$ 13.54 \\

\hline \\
\vspace*{+0.1cm}
Mid Z$^{2}_{2}$  \hfill  $1 - 10$ keV & 1.269 & 42.2~$\%$ | 100$~\%$  & 4961 & 3604 & 38.42 $\pm$ 3.63 & 45.90 & 23.91 & 4.14 $\times 10 ^{-9}$ & 15.43 $\pm$ 1.69 \\
\vspace*{+0.1cm}
$1.21 - 1.31$ Hz & 1.286 & 14.8~$\%$ | 0.00$~\%$  & -- & 0 & 26.84 $\pm$ 1.33 & 31.18 & 23.53 & -2.27 $\times 10 ^{-9}$ & $\leq$ 16.38 \\
\vspace*{+0.3cm}
$n = 2500$ \hfill $R = 12500$ & 1.293 & 15.6~$\%$ | 0.00$~\%$  & -- & 0 & 26.80 $\pm$ 1.43 & 32.76 & 23.01 & 1.21 $\times 10 ^{-9}$ & $\leq$ 18.05 \\

\hline \\
\vspace*{+0.1cm}
Mid Z$^{2}_{3}$  \hfill  $1 - 10$ keV  \\
\vspace*{+0.1cm}
$1.21 - 1.31$ Hz & 1.269 & 85.3~$\%$ | 99.5$~\%$  & 1518 & 61 &  36.88 $\pm$ 2.66 & 47.12 & 30.52 & 4.13 $\times 10 ^{-9}$ & 16.33 $\pm$ 2.00 \\
\vspace*{+0.3cm}
$n = 2500$ \hfill $R = 12500$ \\

\hline \\
\vspace*{+0.1cm}
Fig.~\ref{fig:DetSig} \ Z$^{2}_{3}$  \hfill  $1 - 10$ keV  \\
\vspace*{+0.1cm}
$1.20 - 1.30$ Hz & 1.269 & 71.8~$\%$ | 100$~\%$  & 1175 & 166 &  37.41 $\pm$ 4.04 & 50.55 & 28.79 & 4.40$\times 10 ^{-9}$ & 16.35 $\pm$ 2.00 \\
\vspace*{+0.3cm}
$n = 1000$ \hfill $R = 5000$ \\

\hline \\
\vspace*{+0.1cm}
Fig.~\ref{fig:DetSig} \ Z$^{2}_{3}$  \hfill  $1 - 10$ keV & 1.269 & 62.6~$\%$ | 100$~\%$  & 888 & 63 &  36.77 $\pm$ 3.30 & 49.96 & 28.99 & 3.68$\times 10 ^{-9}$ & 16.32 $\pm$ 2.01 \\
\vspace*{+0.1cm}
$1.21 - 1.31$ Hz & 1.293 & 13.5~$\%$ | 0.00$~\%$  & -- & 0 &  34.90 $\pm$ 1.69 & 40.09 & 29.34 & 8.08$\times 10 ^{-9}$ & $\leq$ 18.44 \\
\vspace*{+0.3cm}
$n = 1000$ \hfill $R = 5000$ \\

\hline \\
\vspace*{+0.1cm}
Narrow Z$^{2}_{3}$  \hfill  $1 - 10$ keV  \\
\vspace*{+0.1cm}
$1.26 - 1.28$ Hz& 1.269 & 91.4~$\%$ | 100$~\%$  & 5164 & 2538 &  36.20 $\pm$ 4.80 & 50.42 & 26.43 & 1.58 $\times 10 ^{-9}$ & 15.98 $\pm$ 2.02 \\
\vspace*{+0.3cm}
$n = 2500$ \hfill $R = 12500$ \\

\hline \\
\vspace*{+0.1cm}
& 0.172 & 1.94~$\%$ | 0.00$~\%$  & -- & -- & 37.40 $\pm$ 3.70 & 47.27 & 31.60 & 4.24 $\times 10 ^{-8}$ & -- \\
\vspace*{+0.1cm}
 & 3.440 & 3.96~$\%$ | 5.94$~\%$  & -- & -- & 37.18 $\pm$ 4.27 & 50.87 & 31.60 & -8.33 $\times 10^{-8}$ & --  \\
\vspace*{+0.1cm}
& 4.594 & 2.62~$\%$ | 8.91 $~\%$  & -- & -- & 37.13 $\pm$ 4.21 & 52.11 & 37.13 & -3.76 $\times 10 ^{-8}$ & -- \\ 
\vspace*{+0.1cm}
Acceleration with & 3.485 & 1.64~$\%$ | 22.9$~\%$  & -- & -- & 37.00 $\pm$ 4.80 & 52.59 & 31.61 & 7.42 $\times 10 ^{-8}$ & -- \\
\vspace*{+0.1cm}
Interbinning & 0.146 & 1.97~$\%$ | 0.00 $~\%$ & -- & -- & 36.04 $\pm$ 3.55 & 48.32 & 36.04 & 6.27 $\times 10 ^{-9}$ & -- \\
\vspace*{+0.1cm}
$0.01 - 5$ Hz & 4.332 & 1.45~$\%$ | 51.5$~\%$  & -- & -- & 35.49 $\pm$ 2.95 & 45.20 & 31.59 & 1.38 $\times 10 ^{-8}$ & -- \\
\vspace*{+0.1cm}
$ 0.2 - 10$ keV & 0.824 & 1.30~$\%$ | 0.00$~\%$ & -- & -- & 35.31 $\pm$ 2.63 & 4.49 & 31.59 & -1.88 $\times 10 ^{-7}$ & -- \\
\vspace*{+0.1cm}
$n = 100$ \hfill $R = 40852$ & 1.822 & 1.02~$\%$ | 0.00$~\%$  & -- & -- & 34.88 $\pm$ 1.96 & 41.29 & 31.61 & -1.72 $\times 10 ^{-7}$ & --  \\
\vspace*{+0.1cm}
& 4.943 & 1.89~$\%$ | 0.00$~\%$ & -- & -- & 34.80 $\pm$ 2.13 & 41.03 & 31.59 & 1.65 $\times 10 ^{-7}$ & -- \\
\vspace*{+0.3cm}
& 2.670 & 2.83~$\%$ | 0.00$~\%$  & -- & -- & 34.13 $\pm$ 2.49 & 57.06 & 31.59 & 8.12 $\times 10 ^{-9}$ & -- \\

\hline \\
\vspace*{+0.1cm}
& 0.172 & 13.5~$\%$ | 98.0$~\%$ & -- & -- &  34.35 $\pm$ 2.60 & 39.61 & 29.16 & 4.18 $\times 10 ^{-8}$ & -- \\
\vspace*{+0.1cm}
Standard Acceleration & 4.165 & 6.44~$\%$ | 0.01$~\%$ & -- & -- & 31.51 $\pm$ 1.77 & 36.27 & 28.67 & 2.07 $\times 10 ^{-7}$ & -- \\
\vspace*{+0.1cm} 
$0.01 - 5$ Hz & 1.838 & 14.2~$\%$ | 0.00$~\%$  & -- & -- & 30.64 $\pm$ 1.23 & 33.73 & 28.52 & -4.03 $\times 10 ^{-8}$ & --\\
\vspace*{+0.1cm} 
 $ 0.2 - 10$ keV & 3.440 & 14.5~$\%$ | 0.01$~\%$  & -- & -- & 30.61 $\pm$ 1.49 & 35.744 & 28.59 & -4.90 $\times 10 ^{-8}$ & --\\

\caption[]{Continued}\\
\vspace*{+0.1cm}
 $n = 100$ \ $R = 1491$ & 0.824 & 14.1~$\%$ | 0.00$~\%$  & -- & -- & 30.59 $\pm$ 1.38 & 34.49 & 28.57 & -1.88 $\times 10 ^{-7}$ & --\\
\vspace*{+0.3cm}
& 1.967 & 4.90~$\%$ | 0.00$~\%$  & -- & -- & 29.78 $\pm$ 0.88 & 32.642 & 28.53 & 1.73 $\times 10 ^{-7}$ & --\\

\hline \\
\vspace*{+0.1cm}
Wide Z$^{2}_{3}$ \hfill $0.2 - 10$ keV & 1.670 & 39.2 ~$\%$ | 85.1$~\%$  & - & 0 & 41.76 $\pm$ 2.58 & 47.19 & 34.61 & -6.83 $\times 10^{-9}$ & $\leq 14.66$ \\
\vspace*{+0.1cm} 
$0.77 - 1.77$ Hz & 1.633 & 5.54 ~$\%$ | 5.94$~\%$  & -- & 0 & 40.16 $\pm$ 2.02 & 42.89 & 35.03 & -3.35 $\times 10 ^{-8}$ & $\leq 14.45$ \\
\vspace*{+0.3 cm} 
$n = 100$ \hfill $R = 500$ & 1.368 & 22.4 ~$\%$ | 7.92$~\%$  & -- & 0 & 39.02 $\pm$ 2.05 & 43.80 & 34.64 & -3.15 $\times 10 ^{-8}$ & $\leq 14.29$\\

\hline \\
\vspace*{+0.1cm}
Mid Z$^{2}_{1}$  \hfill  $0.2 - 10$ keV & 1.256 & 39.6~$\%$ | 50.5$~\%$  & -- & 0 & 17.77 $\pm$ 1.04 & 21.29 & 15.87 & -1.45 $\times 10 ^{-8}$ & $\leq$ 10.07 \\
\vspace*{+0.1cm}
$1.21 - 1.31$ Hz & 1.251 & 18.4~$\%$ | 24.8$~\%$  & -- & 0 & 17.30 $\pm$ 0.97 & 19.58 & 15.01 & 3.37 $\times 10^{-8}$ & $\leq$ 10.01 \\
\vspace*{+0.3cm}
$n = 100$ \hfill $R = 500$ & 1.277 & 16.6~$\%$ | 18.8$~\%$  & -- & 0 & 17.27 $\pm$ 1.03 & 20.55 & 15.47 & -3.37 $\times 10 ^{-8}$ & $\leq$ 10.03\\

\hline \\
\vspace*{+0.1cm}
Mid Z$^{2}_{2}$  \hfill  $0.2 - 10$ keV & 1.270 & 55.4~$\%$ | 100$~\%$  & -- & 0 & 26.66 $\pm$ 2.59 & 34.47 & 22.03 & -3.99 $\times 10 ^{-9}$ & $\leq$ 11.96 \\
\vspace*{+0.1cm}
$1.21 - 1.31$ Hz & 1.303 & 22.4~$\%$ | 0.00$~\%$  & -- & 0 & 24.64 $\pm$ 1.10 & 27.91 & 22.11 & 7.99 $\times 10^{-9}$ & $\leq$ 11.97\\
\vspace*{+0.3cm}
$n = 100$ \hfill $R = 500$ \\

\hline \\
\vspace*{+0.1cm}
Mid Z$^{2}_{3}$  \hfill  $0.2 - 10$ keV & 1.270 & 23.0~$\%$ | 62.4$~\%$  & -- & 0 & 32.94 $\pm$ 2.10 & 38.05 & 28.46 & 6.90 $\times 10 ^{-9}$ & $\leq$ 13.20 \\
\vspace*{+0.1cm}
$1.21 - 1.31$ Hz & 1.303 & 20.0~$\%$ | 25.7$~\%$  & -- & 0 & 31.97 $\pm$ 1.70 & 35.71 & 28.53 & 1.00 $\times 10^{-8}$ & $\leq$ 13.17 \\
\vspace*{+0.1cm}
$n = 100$ \hfill $R = 500$ & 1.275 & 14.7~$\%$ | 3.96$~\%$  & -- & 0 & 30.70 $\pm$ 1.31 & 34.32 & 28.33 & -1.91 $\times 10 ^{-8}$ & $\leq$ 13.25 \\
\vspace*{+0.3cm}
& 1.296 & 14.7~$\%$ | 3.96$~\%$  & -- & 0 & 30.84 $\pm$ 1.48 & 35.39 & 28.44 & -2.49 $\times 10 ^{-8}$ & $\leq$ 13.13 \\

\hline \\
\vspace*{+0.1cm}
Narrow Z$^{2}_{3}$ \hfill $n = 100$\\
\vspace*{+0.1cm}
$R = 500$ \hfill $0.2 - 10$ keV & 1.270 & 70.1~$\%$ | 100$~\%$  & -- & 0 &  28.74 $\pm$ 1.88 & 35.38 & 25.38 & 4.33 $\times 10 ^{-9}$ & $\leq$ 12.78 \\
\vspace*{+0.3cm}
$1.26 - 1.28$ Hz\\

\hline \\
\vspace*{+0.1cm}
Harmonic Z$^{2}_{1}$ & 2.548 & 17.0 ~$\%$ | 28.7$~\%$  & -- & 0 & 19.18 $\pm$ 1.01 & 21.87 & 17.16 & -2.36 $\times 10 ^{-8}$ & $\leq 13.97$ \\
\vspace*{+0.1cm} 
$1 - 10$ keV  & 2.538 & 30.3 ~$\%$ | 16.8$~\%$  & -- & 0 & 18.95 $\pm$ 1.05 & 21.98 & 16.67 & -3.17 $\times 10^{-8}$ & $\leq$ 13.92 \\
\vspace*{+0.1 cm} 
$2.413 - 2.663$ Hz & 2.603 & 11.3 ~$\%$ | 10.9$~\%$  & -- & 0 & 18.85 $\pm$ 1.06 & 21.72 & 16.89 & -3.37 $\times 10 ^{-8}$ & $\leq 13.97$\\
\vspace*{+0.3 cm} 
$n = 100$ \hfill $R = 500$ & 2.440 & 13.3 ~$\%$ | 8.91$~\%$  & -- & 0 & 18.82 $\pm$ 0.91 & 21.19 & 17.07 & 1.72 $\times 10 ^{-8}$ & $\leq 13.96$\\

\hline \\
\vspace*{+0.1cm}
Harmonic Z$^{2}_{1}$ & 3.807 & 41.6 ~$\%$ | 56.4$~\%$  & -- & 0 & 19.55 $\pm$ 1.35 & 24.12 & 16.47 & -20.7 $\times 10 ^{-8}$ & $\leq 13.95$ \\
\vspace*{+0.1cm} 
$1 - 10$ keV  &  8.844 & 15.4~$\%$ | 17.8$~\%$  & -- & 0 & 18.96 $\pm$ 1.52 & 22.16 & 16.49 & -4.82 $\times 10 ^{-9}$ & $\leq 13.99$\\
\vspace*{+0.1 cm} 
$3.682 - 3.932$ Hz &3.775 & 21.6 ~$\%$ | 23.7$~\%$  & -- & 0 & 18.80 $\pm$ 1.22 & 23.00 & 16.65 & -1.78 $\times 10^{-8}$ & $\leq$ 13.97 \\
\vspace*{+0.3 cm} 
$n = 100$ \hfill $R = 500$ \\
\hline 

\end{longtable}
\end{footnotesize}





\bsp	
\label{lastpage}
\end{document}